\documentclass{article}
\usepackage[a4paper,top=3cm,bottom=3cm,left=3cm,right=3cm]{geometry}

\usepackage{oldlfont,amssymb,epsf,stmaryrd,epsfig,color}

\def\lra{\longrightarrow}
\def\ra{\rightarrow}
\def\fa{\forall}
\def\ex{\exists}

\newbox\tempa
\newbox\tempb
\newdimen\tempc
\def\mud#1{\hfil $\displaystyle{\mathstrut #1}$\hfil}
\def\rig#1{\hfil $\displaystyle{#1}$}
\def\irulehelp#1#2#3{\setbox\tempa=\hbox{$\displaystyle{\mathstrut #2}$}%
                        \setbox\tempb=\vbox{\halign{##\cr
        \mud{#1}\cr
        \noalign{\vskip\the\lineskip}
        \noalign{\hrule height 0pt}
        \rig{\vbox to 0pt{\vss\hbox to 0pt{${\; #3}$\hss}\vss}}\cr
        \noalign{\hrule}
        \noalign{\vskip\the\lineskip}
        \mud{\copy\tempa}\cr}}
                      \tempc=\wd\tempb
                      \advance\tempc by \wd\tempa
                      \divide\tempc by 2 }
\def\irule#1#2#3{{\irulehelp{#1}{#2}{#3}
                     \hbox to \wd\tempa{\hss \box\tempb \hss}}}

\newtheorem{definition}{Definition}[section]
\newtheorem{theorem}{Theorem}[section]
\newtheorem{lemma}{Lemma}[section]
\newtheorem{corollary}{Corollary}[section]

\newenvironment{proof}{\noindent {\em Proof.}}{\medskip}
\newenvironment{example}{\noindent {\em Example.}}{\medskip}
\newenvironment{remark}{\noindent {\em Remark.}}{\medskip}

\title{Models and termination of proof reduction in the 
$\lambda \Pi$-calculus modulo theory}

\author{Gilles Dowek\thanks{Inria and 
\'Ecole normale sup\'erieure de Paris-Saclay,
61, avenue du Pr\'esident Wilson,
94235 Cachan Cedex, France,
{\tt gilles.dowek@ens-paris-saclay.fr}}}

\date{}

\begin{document}
\maketitle
\thispagestyle{empty}

\begin{abstract}
We define a notion of model for the $\lambda \Pi$-calculus modulo
theory and prove a soundness theorem. We then define a notion of
super-consistency and prove that proof reduction terminates in the
$\lambda \Pi$-calculus modulo any super-consistent theory. We prove
this way the termination of proof reduction in several theories
including Simple type theory and the Calculus of constructions.
\end{abstract}

\section{Introduction}

\subsection{Models and termination}

In Predicate logic, a model is defined by a domain ${\cal M}$, a set
${\cal B}$ of truth values, and an interpretation function,
parametrized by a valuation $\phi$, mapping each term $t$ to an
element $\llbracket t \rrbracket_{\phi}$ of ${\cal M}$, and each
proposition $A$ to an element $\llbracket A \rrbracket_{\phi}$ of
${\cal B}$.

Predicate logic can be extended to Deduction modulo theory
\cite{DHK,DW}, where a congruence on propositions defining a
computational equality, also known as definitional equality in
Constructive type theory \cite{MartinLof}, is added. Proofs of a
proposition $A$ are then considered to also be proofs of any proposition
congruent to $A$. In Deduction modulo theory, like in Predicate logic, 
a model is defined by a domain ${\cal M}$, a set ${\cal B}$ of truth
values, and an interpretation function.

Usually, the set ${\cal B}$ is the two-element set $\{0,1\}$, but the
notion of model can be extended to a notion of many-valued model, where ${\cal
  B}$ is an arbitrary Boolean algebra, a Heyting algebra, a
pre-Boolean algebra \cite{BHH}, or a pre-Heyting algebra \cite{TVA}.
Boolean algebras permit to introduce intermediate truth values for
propositions that are neither provable nor disprovable, Heyting
algebras to construct models of constructive logic,
and pre-Boolean
and pre-Heyting algebras, where the order relation $\leq$ is replaced
by a pre-order relation, to distinguish a notion of weak
equivalence: $\llbracket A \rrbracket_{\phi} \leq \llbracket B
\rrbracket_{\phi}$ and $\llbracket B \rrbracket_{\phi} \leq \llbracket
A \rrbracket_{\phi}$, for all $\phi$, from a notion of strong
equivalence: $\llbracket A \rrbracket_{\phi} = \llbracket B
\rrbracket_{\phi}$, for all $\phi$. In Deduction modulo theory,
the first corresponds to the
provability of $A \Leftrightarrow B$ and the second to the congruence.

In a model valued in a Boolean algebra, a Heyting algebra, a
pre-Boolean algebra, or a pre-Heyting algebra, a proposition $A$ is
said to be valid when it is weakly equivalent to the proposition
$\top$, that is when, for all $\phi$, $\llbracket A \rrbracket_{\phi}
\geq \tilde{\top}$, and this condition can be rephrased as $\llbracket
A \rrbracket_{\phi} = \tilde{\top}$ in Boolean and Heyting algebras. A
congruence $\equiv$ defined on propositions is said to be valid when,
for all $A$ and $B$ such that $A \equiv B$, $A$ and $B$ are strongly
equivalent, that is, for all $\phi$, 
$\llbracket A \rrbracket_{\phi} = \llbracket B
\rrbracket_{\phi}$.  Note that the relation
$\leq$ is used in the definition of the validity of a proposition, but
not in the definition of the validity of a congruence.

Proof reduction terminates in Deduction modulo a theory defined by a
set of axioms ${\cal T}$ and a congruence $\equiv$, when this theory
has a model valued in the pre-Heyting algebra of reducibility
candidates \cite{DW}. As a consequence, proof reduction terminates
if the theory is super-consistent, that is if, for all pre-Heyting
algebras ${\cal B}$, it has a model valued in ${\cal B}$
\cite{TVA}. This theorem
permits to completely separate the semantic and the syntactic aspects
that are often mixed in the usual proofs of termination of proof
reduction.  The semantic aspect is in the proof of super-consistency
of the considered theory and the syntactic in the universal proof that
super-consistency implies termination of proof reduction.

For the termination of proof reduction, the congruence matters, but
the axioms do not. Thus, the pre-order relation $\leq$ does not matter
in the algebra of reducibility candidates and it is possible to define
it as the trivial pre-order relation such that $C \leq C'$, for all
$C$ and $C'$.  Such a pre-Heyting algebra is said to be trivial.  As
the pre-order is trivial, all the conditions defining pre-Heyting
algebras, such as $a~\tilde{\wedge}~b \leq a$, $a~\tilde{\wedge}~b
\leq b$... are always satisfied in a trivial pre-Heyting algebra,
and a trivial pre-Heyting algebra is just a set equipped with
arbitrary operations $\tilde{\wedge}$, $\tilde{\Rightarrow}$...
Thus, in order to prove that proof reduction terminates in Deduction
modulo a theory defined by a set of axioms ${\cal T}$ and a congruence
$\equiv$, it is sufficient to prove that for all trivial pre-Heyting
algebras ${\cal B}$, the theory has a model valued in ${\cal B}$.

\subsection{The $\lambda \Pi$-calculus modulo theory}

In Predicate logic and in Deduction modulo theory, terms,
propositions, and proofs belong to three distinct languages.  But, it
is more thrifty to consider a single language, such as the $\lambda
\Pi$-calculus modulo theory \cite{CD}, which is implemented in the
\textsc{Dedukti} system \cite{expressing}, or Martin-L\"of's Logical
Framework \cite{NPS}, and express terms, propositions, and proofs, in
this language.
For instance, in Predicate logic, $0$ is a term,
$P(0) \Rightarrow P(0)$ is a proposition and $\lambda
\alpha:P(0)~\alpha$ is a proof of this proposition. In the $\lambda
\Pi$-calculus modulo theory, all these expressions are terms of the
calculus. Only their types differ: $0$ has type $nat$, $P(0)
\Rightarrow P(0)$ has type $Type$ and $\lambda \alpha:P(0)~\alpha$ has
type $P(0) \Rightarrow P(0)$.

Like the $\lambda \Pi$-calculus, the $\lambda \Pi$-calculus modulo
theory is a $\lambda$-calculus with dependent types, but, like in
Deduction modulo theory, its conversion rule is extended to an
arbitrary congruence, typically defined with a confluent and
terminating rewrite system. This idea of extending the conversion rule
beyond $\beta$-reduction is already present in Martin-L\"of type
theory.  It is used, in various ways, in different systems
\cite{NguyenKirchnerKirchner,CirsteaLiquoriWack,FosterStruth,Baueretal}.

\subsection{From pre-Heyting algebras to $\Pi$-algebras}

The first goal of this paper is to extend the notion of pre-Heyting
algebra to a notion of $\Pi$-algebra, adapted to the $\lambda
\Pi$-calculus modulo theory.  

In Predicate logic and in Deduction modulo theory, the propositions
are built from atomic propositions with the connectors and quantifiers
$\top$, $\bot$, $\wedge$, $\vee$, $\Rightarrow$, $\fa$, and $\ex$. 
Accordingly, the operations of a pre-Heyting algebra are
$\tilde{\top}$, $\tilde{\bot}$, $\tilde{\wedge}$, $\tilde{\vee}$,
$\tilde{\Rightarrow}$, $\tilde{\fa}$, and $\tilde{\ex}$. 
In the $\lambda \Pi$-calculus and in the $\lambda \Pi$-calculus 
modulo theory, the only connector is $\Pi$. Thus, a $\Pi$-algebra mainly has
an operation $\tilde{\Pi}$. As expected, its properties are a mixture
of the properties of the implication and of the universal quantifier
of the pre-Heyting algebras.

\subsection{Layered models}

The second goal of this paper is to extend the usual notion of model to
the $\lambda \Pi$-calculus modulo theory.

Extending the notion of model to many-sorted predicate logic requires
to consider not just one domain ${\cal M}$, but a family of domains
${\cal M}_s$ indexed by the sorts. For instance, in a model of Simple
type theory, the family of domains is indexed by simple types.
In the $\lambda\Pi$-calculus modulo theory, the sorts also are just
terms of the calculus.  Thus, we shall define a model of the $\lambda
\Pi$-calculus modulo theory by a family of domains $({\cal M}_t)_t$
indexed by the terms of the calculus and a function $\llbracket
. \rrbracket$ mapping each term $t$ of type $A$ and valuation $\phi$
to an element $\llbracket t \rrbracket_{\phi}$ of ${\cal M}_A$.

The functions ${\cal M}$ and $\llbracket . \rrbracket$ are 
similar, in the sense that both their domains is the set of terms of
the calculus.  The goal of the model construction is to define the
function $\llbracket . \rrbracket$ and the function ${\cal M}$ can be
seen as a tool helping to define this function.  For instance, if
$f$ is a constant of type $A \ra A$, where $A$ is a term of type $Type$,
and we have the rule 
$$f(x) \lra x$$
we want to define the interpretation 
$\llbracket f
\rrbracket$ as the identity function over some set, but to state
which, we must first define the function ${\cal M}$ that maps the term
$A$ to a set ${\cal M}_A$, and then define $\llbracket f \rrbracket$
as the identity function over the set ${\cal M}_A$.

In Predicate logic and in Deduction modulo theory, terms 
may be typed with sorts, 
but the sorts themselves have no type. In the $\lambda \Pi$-calculus modulo 
theory, in contrast, terms have types that have types...  
This explains that, in some cases, constructing the function ${\cal M}$
itself requires to define first another function ${\cal N}$, 
that is used as a tool helping to define this function. This can be
iterated to a several layer model, where the function
$\llbracket . \rrbracket$ is defined with the help of a function
${\cal M}$, that is defined with the help of a function ${\cal N}$,
that is defined with the help... 
The number of layers depends on the model.
Such layered constructions are
common in proofs of termination of proof reduction
\cite{Geuvers,MelliesWerner,Blanqui}, for instance for Pure Type 
Systems where sorts are stacked: $Type_0:Type_1:Type_2:Type_3$.

Note that, in this definition of the notion of model, when a term $t$
has type $A$, we do not require $\llbracket t \rrbracket_{\phi}$ to be
an element of $\llbracket A \rrbracket_{\phi}$, but of ${\cal M}_A$.
This is consistent with the notion of model of many-sorted predicate
logic, where we require $\llbracket t \rrbracket_{\phi}$ to be an
element of ${\cal M}_s$ and where $\llbracket s \rrbracket_{\phi}$ is
often not even defined.

Valuations must be handled with care in such layered models. 
In a three layer model, for instance, the definition
of ${\cal N}_t$ is absolute, the definition of 
${\cal M}_t$ is relative to a valuation $\psi$, mapping each 
variable of type $A$ to an element of ${\cal N}_A$, and 
the definition of $\llbracket t \rrbracket$ is relative to 
a valuation $\psi$ and to a valuation $\phi$ mapping each variable 
of type $A$ to an element of ${\cal M}_{A, \psi}$.

\subsection{Super-consistency and proof reduction}

The third goal of this paper is to use this notion of $\Pi$-algebra to
define a notion of super-consistency and to prove that proof
reduction, that is $\beta$-reduction,
 terminates in the $\lambda \Pi$-calculus modulo any
super-consistent theory.

We prove this way the termination of proof reduction in several
theories expressed in the $\lambda \Pi$-calculus modulo theory,
including Simple type theory \cite{DHK} and the Calculus of
constructions \cite{CD}. Together with confluence, this termination of
proof reduction is a property required to define these theories
in  the system \textsc{Dedukti} \cite{expressing}. 

\medskip

In Section \ref{sec:lpm}, we recall the definition of the
$\lambda \Pi$-calculus modulo theory and give three examples of
theories expressed in this framework. In Section \ref{sec:models}, we
introduce the notion of $\Pi$-algebra and that of model for the
$\lambda \Pi$-calculus modulo theory and we prove a soundness theorem.
In Section \ref{sec:sc}, we define the notion of super-consistency and
prove that the three theories introduced in Section \ref{sec:lpm} are
super-consistent. In Section \ref{sec:term}, we prove that proof
reduction terminates in the $\lambda \Pi$-calculus modulo any
super-consistent theory.

\section{The $\lambda \Pi$-calculus modulo theory}
\label{sec:lpm}

\subsection{The $\lambda \Pi$-calculus}

The syntax of the $\lambda \Pi$-calculus is 
$$t = x~|~Type~|~Kind~|~\Pi x:t~t~|~\lambda x:t~t~|~t~t$$
and the typing rules are given in Figure \ref{fig:lp}.

\begin{figure*}[t!]
\framebox{\parbox{\textwidth}{
$$
\hspace{-5cm}
\begin{array}{c}
\irule{} 
      {[~]~\mbox{well-formed}}
      {\mbox{\bf Empty}}\\
\irule{\Gamma \vdash A:Type} 
      {\Gamma, x:A~~\mbox{well-formed}}
      {x \not\in \Gamma~\mbox{\bf Declaration (for object variables)}}\\
\irule{\Gamma \vdash A:Kind} 
      {\Gamma, x:A~~\mbox{well-formed}}
      {x \not\in \Gamma~\mbox{\bf Declaration (for type family variables)}}\\
\irule{\Gamma~\mbox{well-formed}} 
      {\Gamma \vdash Type:Kind}
      {\mbox{\bf Sort}}\\
\irule{\Gamma~\mbox{well-formed}} 
      {\Gamma \vdash x:A}
      {x:A \in \Gamma~\mbox{\bf Variable}}\\
\irule{\Gamma \vdash A:Type~~\Gamma, x:A \vdash B:Type}
      {\Gamma \vdash \Pi x:A~B:Type}
      {\mbox{\bf Product (for types)}}\\
\irule{\Gamma \vdash A:Type~~\Gamma, x:A \vdash B:Kind}
      {\Gamma \vdash \Pi x:A~B:Kind}
      {\mbox{\bf Product (for kinds)}}\\
\irule{\Gamma \vdash A:Type~~\Gamma, x:A \vdash B:Type~~\Gamma, x:A \vdash t:B} 
      {\Gamma \vdash \lambda x:A~t:\Pi x:A~B} 
      {\mbox{\bf Abstraction (for objects)}}\\
\irule{\Gamma \vdash A:Type~~\Gamma, x:A \vdash B:Kind~~\Gamma, x:A \vdash t:B} 
      {\Gamma \vdash \lambda x:A~t:\Pi x:A~B} 
      {\mbox{\bf Abstraction (for type families)}}\\
\irule{\Gamma \vdash t:\Pi x:A~B~~\Gamma \vdash u:A} 
      {\Gamma \vdash (t~u):(u/x)B} 
      {\mbox{\bf Application}}\\
\irule{\Gamma \vdash A:Type~~\Gamma \vdash B:Type~~\Gamma \vdash t:A}
      {\Gamma \vdash t:B}
      {A \equiv_{\beta} B~\mbox{\bf Conversion (for types)}}\\
\irule{\Gamma \vdash A:Kind~~\Gamma \vdash B:Kind~~\Gamma \vdash t:A}
      {\Gamma \vdash t:B}
      {A \equiv_{\beta} B~\mbox{\bf Conversion (for kinds)}}
\end{array}$$
\caption{The $\lambda \Pi$-calculus \label{fig:lp}}}}
\end{figure*}

As usual, we write $A \ra B$ for $\Pi x:A~B$ when $x$ does not occur
in $B$.
The $\alpha$-equivalence relation is defined as usual and terms are
identified modulo $\alpha$-equivalence.  The relation $\beta$---one
step $\beta$-reduction at the root---is defined as usual. If
$r$ is a relation on terms, we write $\lra_r^1$ for the congruence
closure of $r$, $\lra_r^+$ for the transitive closure of 
$\lra_r^1$, $\lra_r^*$ for its reflexive-transitive closure,
and $\equiv_r$ for its reflexive-symmetric-transitive closure.

If $\Sigma$, $\Gamma$, and $\Delta$ are contexts, a substitution
$\theta$, binding the variables of $\Gamma$, is said to {\em have
type $\Gamma \leadsto \Delta$ in $\Sigma$} if for all $x:A$ in $\Gamma$, 
we have $\Sigma, \Delta \vdash
\theta x:\theta A$.  In this case, if $\Sigma, \Gamma \vdash t:B$,
then $\Sigma, \Delta \vdash \theta t:\theta B$.

Types are preserved by $\beta$-reduction. The $\beta$-reduction
relation is confluent and strongly terminating. And each term has a
unique type modulo $\beta$-equivalence \cite{HHP}.

A term $t$, well-typed in some context $\Gamma$, is a {\em
  kind} if its type in this context is $Kind$. For instance, $Type$
and $nat \ra Type$ are kinds. It is a {\em type family} if
its type is a kind. In particular, it is a {\em type} if
its type is $Type$. For instance, $nat$, $array$, and $(array~0)$ are
type families, among which $nat$ and $(array~0)$ are types.  It is 
an {\em object} if its type is a type. For instance, $0$ and
$[0]$ are objects.

\subsection{The $\lambda \Pi$-calculus modulo theory}

\begin{definition}[Rewrite rule]
A {\em rewrite rule} is a triple $l \lra^{\Gamma} r$ where
$\Gamma$ is a context and $l$ and $r$ are $\beta$-normal terms.
Such a rule is {\em well-typed} in the context $\Sigma$ if,
in the $\lambda \Pi$-calculus, the context $\Sigma, \Gamma$ is
well-formed and there exists a term $A$ such that the terms $l$ and $r$ both 
have type $A$ in this context.
\end{definition}

If $\Sigma$ is a context, $l \lra^{\Gamma} r$ is a rewrite rule
well-typed in $\Sigma$ and $\theta$ is a substitution of type $\Gamma
\leadsto \Delta$ in $\Sigma$, then the terms $\theta l$ and $\theta r$
both have type $\theta A$ in the context $\Sigma, \Delta$.
The relation ${\cal R}$---one step ${\cal R}$-reduction at the root---is
defined by: $t~{\cal R}~u$ is there exists a rewrite rule $l
\lra^{\Gamma} r$ and a substitution $\theta$ such that $t = \theta
l$ and $u = \theta r$.  The relation $\beta {\cal R}$---one step $\beta
{\cal R}$-reduction at the root---is the union of $\beta$ and ${\cal
  R}$.

\medskip

\begin{definition}[Theory]
A {\em theory} is a pair formed with a context $\Sigma$, well-formed
in the $\lambda \Pi$-calculus, and a set of rewrite rules ${\cal R}$,
well-typed in $\Sigma$ in the $\lambda \Pi$-calculus.
\end{definition}

The variables declared in $\Sigma$ are called {\em constants}.  
They replace the sorts, the function symbols,
the predicate symbols, and the axioms of a theory in Predicate logic.

\medskip

\begin{definition}[The $\lambda \Pi$-calculus modulo theory]
The {\em $\lambda \Pi$-calculus modulo $\Sigma, {\cal R}$} is the
extension of the $\lambda \Pi$-calculus obtained modifying the {\bf
Declaration} rules to replace the condition $x \not\in \Gamma$ with
$x \not\in \Sigma, \Gamma$, the {\bf Variable} rules to replace the
condition $x:A \in \Gamma$ by $x:A \in \Sigma, \Gamma$, and the {\bf
Conversion} rules to replace the condition $A \equiv_{\beta} B$ with
$A \equiv_{\beta {\cal R}} B$.
\end{definition}

In this paper, we assume that the relation $\lra^1_{\beta {\cal R}}$ 
is confluent and has the subject reduction property.
Confluence and subject reduction are indeed needed to
build models and prove termination of proof reduction.
This is consistent with the
methodology proposed in \cite{Jouannaud}: first prove confluence and 
subject reduction, then termination. 

\subsection{Examples of theories}
\label{sec:Examples}

\begin{figure}[t!]
\framebox{\parbox{\columnwidth}{
$$\begin{array}{rcl}
\iota&:&Type\\
o&:&Type\\
\Rightarrow&:&o \ra o \ra o\\
\fa_A&:&(A \ra o) \ra o\\
\varepsilon&:&o \ra Type\\
\\
(\varepsilon~(\Rightarrow~x~y)) &\lra& (\varepsilon~x) \ra (\varepsilon~y)\\
(\varepsilon~(\fa_A~x)) &\lra& \Pi z:A~(\varepsilon~(x~z))
\end{array}$$
with a finite number of quantifiers $\fa_A$
\caption{Simple type theory \label{fig:stt}}}}
\end{figure}

Simple type theory can be expressed in Deduction modulo theory
\cite{DHKHOL}. The main idea in this presentation is to distinguish
terms of type $o$ from propositions. If $t$ is a term of type $o$, the
corresponding proposition is written $\varepsilon(t)$. The term $t$ is a
{\em propositional content} or a {\em code} of the proposition 
$\varepsilon(t)$. This way, it is not possible
to quantify over propositions, but it is possible to quantify over
codes of propositions: there is no proposition 
$$\fa X~(X \Rightarrow X)$$
but there is a proposition 
$$\fa x~(\varepsilon(x) \Rightarrow \varepsilon(x))$$
respecting the syntax of Predicate logic, where the predicate symbol
$\varepsilon$ is applied to the variable $x$ to form a proposition.

In this presentation, each simple type is a sort and, 
for each simple type $A$, there is a quantifier $\fa_A$.  Thus, the
language contains an infinite number of sorts and an infinite number
of constants.

This presentation can be adapted to the $\lambda \Pi$-calculus modulo
theory.  To avoid declaring an infinite number of constants for simple
types, we can just declare two constants $\iota$ and $o$ of type
$Type$ and use the product of the $\lambda \Pi$-calculus modulo
theory to represent the simple types $\iota \ra \iota$, $\iota \ra
\iota \ra \iota$, $\iota \ra o$... We should declare an infinite number
of quantifiers $\fa_A$, indexed by simple types, but this can be avoided
as, in each specific proof, only a finite number of such quantifiers 
occur. This leads to the theory presented in Figure \ref{fig:stt}.

Another possibility is to add the type $A$ as an extra argument of the
quantifier $\fa$. To do so, we need to introduce a type
$type$ for codes of simple types, two constants $\iota$ and $o$, of
type $type$, and not $Type$, a constant $arrow$ of type $type \ra type
\ra type$, and a decoding function $\eta$ of type $type \ra Type$. 
This way, the quantifier $\fa$ can be
given the type $\Pi a:type~(((\eta~a) \ra (\eta~o)) \ra (\eta~o))$.
This leads to the theory presented in Figure \ref{fig:sttp}.

The Calculus of constructions \cite{CoquandHuet} can also be expressed
in the $\lambda \Pi$-calculus modulo theory \cite{CD} as the theory
presented in Figure \ref{fig:coc}.
Note that this presentation slightly differs from that of \cite{CD}:
the symbol $U_{Type}$ has been replaced everywhere by 
$\varepsilon_{Kind}(\dot{Type})$ allowing to
drop the rule 
$$\varepsilon_{Kind} (\dot{Type}) \lra U_{Type}$$
Then, to keep the notations similar to those of Simple type theory, 
we write 
$type$ for $U_{Kind}$, 
$o$ for $\dot{Type}$, 
$\eta$ for $\varepsilon_{Kind}$, and 
$\varepsilon$ for $\varepsilon_{Type}$.
We also write 
$\dot{\Pi}_{KK}$ for $\dot{\Pi}_{\langle Kind, Kind, Kind \rangle}$, 
$\dot{\Pi}_{TT}$ for $\dot{\Pi}_{\langle Type, Type, Type \rangle}$, 
$\dot{\Pi}_{KT}$ for $\dot{\Pi}_{\langle Kind, Type, Type \rangle}$, 
and $\dot{\Pi}_{TK}$ for $\dot{\Pi}_{\langle Type, Kind, Kind \rangle}$.
Note finally that the symbol $\dot{\Pi}_{KT}$ is exactly the 
parametric universal quantifier of Simple type theory, 
the symbol $\dot{\Pi}_{TT}$
is a dependent version of the symbol $\Rightarrow$ and 
$\dot{\Pi}_{KK}$ a dependent version of the symbol $arrow$. 
The symbol $\dot{\Pi}_{TK}$, in contrast, is new.

\begin{figure}[t!]
\framebox{\parbox{\columnwidth}{
$$
\hspace*{-0.32cm}
\begin{array}{rcl}
type&:&Type\\
\iota&:&type\\
o&:&type\\
arrow&:&type \ra type \ra type\\
\eta&:&type \ra Type\\
\Rightarrow&:&(\eta~o) \ra (\eta~o) \ra (\eta~o)\\
\fa&:&\Pi a:type~(((\eta~a) \ra (\eta~o)) \ra (\eta~o))\\
\varepsilon&:&(\eta~o) \ra Type\\
\\
(\eta~(arrow~x~y)) &\lra& (\eta~x) \ra (\eta~y)\\
(\varepsilon~(\Rightarrow~x~y)) &\lra& (\varepsilon~x) \ra (\varepsilon~y)\\
(\varepsilon~(\fa~x~y)) &\lra& \Pi z:(\eta~x)~(\varepsilon~(y~z))
\end{array}$$
\caption{Simple type theory with a parametric quantifier\label{fig:sttp}}}}
\end{figure}

\section{Algebras and Models}
\label{sec:models}

\subsection{$\Pi$-algebras}

The notion of $\Pi$-algebra is an adaptation 
of that of pre-Heyting algebra to the $\lambda
\Pi$-calculus.

\medskip

\begin{definition}[$\Pi$-algebra]
A $\Pi$-algebra is formed with 
\begin{itemize}
\item a set ${\cal B}$, 
\item a pre-order relation $\leq$ on ${\cal B}$, 
\item an element $\tilde{\top}$ of ${\cal B}$,
\item a function $\tilde{\wedge}$ from ${\cal B} \times {\cal B}$ to ${\cal B}$,
\item a subset ${\cal A}$ of ${\cal P}^{+}({\cal B})$, the set of non-empty 
subsets of ${\cal B}$, 
\item a function $\tilde{\Pi}$ from ${\cal B} \times {\cal A}$ to ${\cal B}$,
\end{itemize}
such that 
\begin{itemize}
\item 
$\tilde{\top}$  is a maximal element for $\leq$, that is for all $a$ in 
${\cal B}$, $a \leq \tilde{\top}$, 
\item 
$a~\tilde{\wedge}~b$ is a greatest lower bound of $\{a,b\}$ for $\leq$,
that is $a~\tilde{\wedge}~b \leq a$, $a~\tilde{\wedge}~b \leq b$,
and for all $c$, if $c \leq a$ and $c \leq b$, then $c \leq
a~\tilde{\wedge}~b$,
\item
$a \leq \tilde{\Pi}(b,S)$ if and only if for all $c$ in $S$, 
$a~\tilde{\wedge}~b \leq c$.
\end{itemize}
\end{definition}

Note that is the relation $\leq$ is a pre-order, and not necessarily 
an order, greatest lower bounds are not necessarily unique, when they exist.

Note also that, from the operation $\tilde{\Pi}$, we can define an
exponentiation operation $b~\tilde{\ra}~c = \tilde{\Pi}(b,\{c\})$ that
verifies the usual properties of exponentiation: $a \leq
b~\tilde{\ra}~c$ if and only if $a~\tilde{\wedge}~b \leq c$.  When the
set $S$ has a greatest lower bound $\tilde{\bigwedge} S$, the
operation mapping $b$ and $S$ to $b~\tilde{\ra}~\tilde{\bigwedge} S$
verifies the same properties as $\tilde{\Pi}$: $a \leq
b~\tilde{\ra}~\tilde{\bigwedge} S$ if and only if $a~\tilde{\wedge}~b
\leq \tilde{\bigwedge} S$ if and only if for all $c$ in $S$,
$a~\tilde{\wedge}~b \leq c$.  But this decomposition is possible only
when all sets of ${\cal A}$ have greatest lower bounds.

\medskip

\begin{definition}[Full $\Pi$-algebra]
A $\Pi$-algebra is {\em full} if ${\cal A} = {\cal P}^{+}({\cal B})$, that is 
if $\tilde{\Pi}$ is total on  ${\cal B} \times {\cal P}^{+}({\cal B})$.
\end{definition}

\medskip

\begin{example}
\label{ex:01}
The algebra
$\langle \{0,1\},1,\tilde{\wedge},{\cal P}^+(\{0,1\}),\tilde{\Pi}
\rangle$,
where $\tilde{\wedge}$ and $\tilde{\Pi}$ are defined by the tables
below, is a $\Pi$-algebra. Note that, dropping the middle column of
the table of $\tilde{\Pi}$, we get the table of implication and,
dropping the first line, that of the universal quantifier.
$$\begin{array}{|c|c|c|}
\hline
\tilde{\wedge}&0&1\\
\hline
0&0&0\\
\hline
1&0&1\\
\hline
\end{array}
\hspace{2cm}
\begin{array}{|c|c|c|c|}
\hline \tilde{\Pi}&\{0\}&\{0,1\}&\{1\}\\
\hline
0&1&1&1\\
\hline
1&0&0&1\\
\hline
\end{array}$$
\end{example}

\begin{figure*}[t!]
\framebox{\parbox{\textwidth}{
$$\begin{array}{rcl}
type&:&Type\\
o&:&type\\
\eta&:&type \ra Type\\
\varepsilon&:&(\eta~o) \ra Type\\
\dot{\Pi}_{KK} &:& \Pi x:type~(((\eta~x) \ra type) \ra type)\\
\dot{\Pi}_{TT} &:& \Pi x:(\eta~o)~(((\varepsilon~x) \ra (\eta~o)) \ra (\eta~o))\\
\dot{\Pi}_{KT} &:& \Pi x:type~(((\eta~x) \ra (\eta~o)) \ra (\eta~o))\\
\dot{\Pi}_{TK} &:& \Pi x:(\eta~o)~(((\varepsilon~x) \ra type) \ra type)\\
\\
(\eta~(\dot{\Pi}_{KK}~x~y)) &\lra&
\Pi z:(\eta~x)~(\eta~(y~z))\\
(\varepsilon~(\dot{\Pi}_{TT}~x~y)) &\lra&
\Pi z:(\varepsilon~x)~(\varepsilon~(y~z))\\
(\varepsilon~(\dot{\Pi}_{KT}~x~y)) &\lra&
\Pi z:(\eta~x)~(\varepsilon~(y~z))\\
(\eta~(\dot{\Pi}_{TK}~x~y)) &\lra&
\Pi z:(\varepsilon~x)~(\eta~(y~z))
\end{array}$$
\caption{The Calculus of constructions \label{fig:coc}}}}
\end{figure*}

\subsection{Models valued in a $\Pi$-algebra ${\cal B}$}

\begin{definition}[Model]
A {\em model} is a family of interpretation functions ${\cal D}^1,
..., {\cal D}^n$ such that for all $i$, ${\cal D}^i$ is a function
mapping each term $t$ of type $B$ in some context $\Gamma$, 
function $\phi_1$ mapping each variable $x:A$ of $\Gamma$ to an element of 
${\cal D}^1_A$, ..., and function $\phi_{i-1}$ mapping each variable
$x:A$ of $\Gamma$ to an element of ${\cal D}^{i-1}_{A, \phi_1, ...,
\phi_{n-2}}$, to some ${\cal D}^i_{t,\phi_1, ..., \phi_{i-1}}$ in
${\cal D}^{i-1}_{B,\phi_1,...,\phi_{i-2}}$, and for all $t$, $u$,
$\phi_1$, ..., $\phi_{n-1}$
$${\cal D}^n_{(u/x)t, \phi_1, ..., \phi_{n-1}} = 
{\cal D}^n_{t, (\phi_1, x = {\cal D}^1_{u}), ..., (\phi_{n-1}, x = {\cal D}^{n-1}_{u, 
\phi_1, ..., \phi_{n-2}})}$$
\end{definition}

For the last function ${\cal D}^n$, we write 
$\llbracket t \rrbracket_{\phi_1, ..., \phi_{n-1}}$ instead of
${\cal D}^n_{t, \phi_1, ..., \phi_{n-1}}$. 

In the examples presented in this paper, we use the cases $n = 2$ and
$n = 3$ only.  The general definition then specializes as follows.

\begin{example}
When $n = 2$, a model is given by two functions
${\cal M}$ and $\llbracket . \rrbracket$ such that
\begin{itemize}
\item ${\cal M}$ is a function mapping each term $t$ of type $B$ in
$\Gamma$ to some ${\cal M}_t$,

\item $\llbracket . \rrbracket$ is a function mapping each term $t$ of
type $B$ in $\Gamma$ and function $\phi$ mapping each
variable $x:A$ of $\Gamma$ to an element of ${\cal M}_A$, to some
$\llbracket t \rrbracket_{\phi}$ in ${\cal M}_B$,
such that for all $t$, $u$ and $\phi$
$$\llbracket (u/x)t \rrbracket_{\phi} = \llbracket t \rrbracket_{\phi,
x = \llbracket u \rrbracket_{\phi}}$$
\end{itemize}
This generalizes of the usual definition of {\em model} for many-sorted 
predicate logic.
\end{example}

\begin{remark}
If $f$ is a constant of type $A \ra A \ra A$, we 
can define the function $\hat{f}$ mapping $a$ and $b$ in ${\cal M}_A$ to 
$\llbracket (f~x~y) \rrbracket_{x = a, y = b}$. Using the property
$\llbracket (u/x)t \rrbracket_{\phi} = \llbracket t \rrbracket_{\phi,
x = \llbracket u \rrbracket_{\phi}}$,
we then get 
$$\llbracket (f~t~u) \rrbracket_{\phi} =
\hat{f}(\llbracket t \rrbracket_{\phi}, \llbracket u \rrbracket_{\phi})$$
which is the usual definition of an interpretation.
\end{remark}

\begin{remark}
The first interpretation function ${\cal M}$ does not depend on any 
valuation, so it must be very rudimentary. For instance in Definition
\ref{MSTT} below, for all objects and most types, we have
${\cal M}_t = \{e\}$. 
Only the types $o$, $o \ra o$... are interpreted in a non trivial 
way. Nevertheless, it is sufficient to support Definition
\ref{bracketSTT}.
\end{remark}

\begin{example}
When $n = 3$, a model is given by three functions
${\cal N}$, ${\cal M}$, and $\llbracket . \rrbracket$ such that
\begin{itemize}
\item ${\cal N}$ is a function mapping each term $t$ of type $B$ in 
$\Gamma$ to some ${\cal N}_t$, 

\item ${\cal M}$ is a function mapping each term $t$ of type $B$ in
$\Gamma$ and function $\psi$ mapping each variable $x:A$
of $\Gamma$ to an element of ${\cal N}_A$, to some ${\cal M}_{t,\psi}$ 
in ${\cal N}_B$, 

\item $\llbracket . \rrbracket$ is a function mapping each term $t$ of
type $B$ in $\Gamma$, function $\psi$ mapping each
variable $x:A$ of $\Gamma$ to an element of ${\cal N}_A$, and
function $\phi$ mapping each variable $x:A$ of $\Gamma$ to
an element of ${\cal M}_{A, \psi}$, to some $\llbracket t \rrbracket_{\psi,\phi}$
in ${\cal M}_{B, \psi}$,
such that for all $t$, $u$, $\psi$, and $\phi$
$$\llbracket (u/x)t \rrbracket_{\psi, \phi} = 
\llbracket t \rrbracket_{(\psi, x = {\cal M}_{u,\psi}), (\phi, x = \llbracket u \rrbracket_{\psi,\phi})}$$
\end{itemize}
\end{example}

\medskip

\begin{definition}[Model valued in a $\Pi$-algebra ${\cal B}$]
Let ${\cal B} = \langle {\cal B}, \tilde{\top}, \tilde{\wedge}, {\cal
A}, \tilde{\Pi} \rangle$ be a $\Pi$-algebra.
A model is {\em valued in ${\cal B}$} if 
\begin{itemize}
\item ${\cal D}^{n-1}_{Kind, \phi_1, ..., \phi_{n-2}} = 
{\cal D}^{n-1}_{Type, \phi_1, ..., \phi_{n-2}} = {\cal B}$, 

\item 
$\llbracket Kind \rrbracket_{\phi_1, ..., \phi_{n-1}} = 
\llbracket Type \rrbracket_{\phi_1, ..., \phi_{n-1}} = \tilde{\top}$

\item 
$\llbracket \Pi x:C~D \rrbracket_{\phi_1,...,\phi_{n-1}}
= \tilde{\Pi}(\llbracket C \rrbracket_{\phi_1, ..., \phi_{n-1}}, 
\{\llbracket D \rrbracket_{(\phi_1, x = c_1), ..., (\phi_{n-1}, x = c_{n-1})}~|~c_1 
\in {\cal D}^1_C, ..., c_{n-1} \in {\cal D}^{n-1}_{C,\phi_1, ..., \phi_{n-2}}\})$
\end{itemize}
\end{definition}

\medskip

We often write $\overline{\phi}$ for a sequence $\phi_1, ..., \phi_n$
and, if $\overline{c} = c_1, ..., c_n$, we write 
$\overline{\phi}, x = \overline{c}$ for the sequence
$(\phi_1, x = c_1), ..., (\phi_n, x = c_n)$. 

\medskip

\begin{definition}[Validity]
A model ${\cal M}$ valued in some $\Pi$-algebra ${\cal B}$ is
{\em model} of a theory $\Sigma, {\cal R}$, or 
the theory is {\em valid} in the model, if 
\begin{itemize}
\item for all constants $c:A$ in $\Sigma$, we have 
$\llbracket A \rrbracket \geq \tilde{\top}$, 
\item and for all $A$ and $B$ well-typed in a context $\Gamma$, 
such that $A \equiv_{\beta {\cal R}} B$, we have 
for all $i$, for all $\overline{\phi}$, 
${\cal D}^i_{A, \overline{\phi}} = {\cal D}^i_{B, \overline{\phi}}$.
\end{itemize}
\end{definition}

\begin{theorem}[Soundness]
Let ${\cal M}$ be a model, valued in some $\Pi$-algebra ${\cal B}$,
of a theory $\Sigma, {\cal R}$.  
Then, for all judgments
$x_1:A_1, ..., x_p:A_p \vdash t:B$ 
derivable in $\Sigma, {\cal R}$, and for all $\overline{\phi}$, we
have
$$\llbracket A_1 \rrbracket_{\overline{\phi}}~\tilde{\wedge}~...~\tilde{\wedge}~
\llbracket A_p \rrbracket_{\overline{\phi}} \leq \llbracket B \rrbracket_{\overline{\phi}}$$
\end{theorem}

\begin{proof}
By induction on the structure of the derivation of 
$x_1:A_1, ..., x_p:A_p \vdash t:B$. 
\begin{itemize}
\item If the last rule is {\bf Sort} or {\bf Product}, then 
$B = Type$ or $B = Kind$, 
$\llbracket B \rrbracket_{\overline{\phi}} = \tilde{\top}$
and 
$$\llbracket A_1 \rrbracket_{\overline{\phi}}~\tilde{\wedge}~...~\tilde{\wedge}~
\llbracket A_p \rrbracket_{\overline{\phi}} \leq \llbracket B \rrbracket_{\overline{\phi}}$$

\item If the last rule is {\bf Variable}, with a constant of $\Sigma$, 
then $\llbracket B \rrbracket_{\overline{\phi}} \geq \tilde{\top}$
and 
$$\llbracket A_1 \rrbracket_{\overline{\phi}}~\tilde{\wedge}~...~\tilde{\wedge}~
\llbracket A_p \rrbracket_{\overline{\phi}} \leq \llbracket B \rrbracket_{\overline{\phi}}$$

\item If the last rule is {\bf Variable}, with a variable of $\Gamma$, 
then $B = A_i$ and
$$\llbracket A_1 \rrbracket_{\overline{\phi}}~\tilde{\wedge}~...~\tilde{\wedge}~
\llbracket A_p \rrbracket_{\overline{\phi}} \leq \llbracket B \rrbracket_{\overline{\phi}}$$

\item If the last rule is {\bf Abstraction}, then $B = \Pi x:C~D$
and by induction hypothesis, for all $\overline{c}$
in ${\cal D}^1_C \times {\cal D}^2_{C,\phi_1} \times ... \times 
{\cal D}^n_{C, \phi_1, ..., \phi_{n-1}}$, 
we have 
$$\llbracket A_1 \rrbracket_{\overline{\phi}}~\tilde{\wedge}~...~\tilde{\wedge}~\llbracket 
A_p \rrbracket_{\overline{\phi}}~\tilde{\wedge}~\llbracket C \rrbracket_{\overline{\phi}} \leq
\llbracket D \rrbracket_{\overline{\phi}, x = \overline{c}}$$
thus 
$$\llbracket A_1 \rrbracket_{\overline{\phi}}~\tilde{\wedge}~...~\tilde{\wedge}~\llbracket 
A_p \rrbracket_{\overline{\phi}}\leq \tilde{\Pi} (\llbracket C \rrbracket_{\overline{\phi}}, 
\{\llbracket D \rrbracket_{\overline{\phi}, x = \overline{c}}
~|~
\overline{c} \in {\cal D}^1_C \times {\cal D}^2_{C,\phi_1} \times ... \times 
{\cal D}^n_{C, \phi_1, ..., \phi_{n-1}}\})$$
$$\llbracket A_1 \rrbracket_{\overline{\phi}}~\tilde{\wedge}~...~\tilde{\wedge}~
\llbracket A_p \rrbracket_{\overline{\phi}} \leq \llbracket \Pi x:C~D \rrbracket_{\overline{\phi}}$$
that is 
$$\llbracket A_1 \rrbracket_{\overline{\phi}}~\tilde{\wedge}~...~\tilde{\wedge}~
\llbracket A_p \rrbracket_{\overline{\phi}} \leq \llbracket B \rrbracket_{\overline{\phi}}$$

\item If the last rule is {\bf Application}, then we have $B = (u/x)D$ and 
by, induction hypothesis
$$\llbracket A_1 \rrbracket_{\overline{\phi}}~\tilde{\wedge}~...~\tilde{\wedge}~
\llbracket A_p \rrbracket_{\overline{\phi}} \leq \llbracket C \rrbracket_{\overline{\phi}}$$
and
$$\llbracket A_1 \rrbracket_{\overline{\phi}}~\tilde{\wedge}~...~\tilde{\wedge}~
\llbracket A_p \rrbracket_{\overline{\phi}} \leq \llbracket \Pi x:C~D \rrbracket_{\overline{\phi}}$$
Thus, for all $\overline{c}$
in ${\cal D}^1_C \times {\cal D}^2_{C,\phi_1} \times ... \times 
{\cal D}^n_{C, \phi_1, ..., \phi_{n-1}}$, we have 
$$\llbracket A_1 \rrbracket_{\overline{\phi}}~\tilde{\wedge}~...~\tilde{\wedge}~
\llbracket A_p \rrbracket_{\overline{\phi}}~\tilde{\wedge}~\llbracket C \rrbracket_{\overline{\phi}} 
\leq \llbracket D \rrbracket_{\overline{\phi}, x = \overline{c}}$$
In particular, for $\overline{c} =
{\cal D}^1_u, {\cal D}^2_{u, \phi_1}, ..., {\cal D}^i_{u, \phi_1, ..., \phi_{i-1}}$, 
we get 
$$\llbracket A_1 \rrbracket_{\overline{\phi}}~\tilde{\wedge}~...~\tilde{\wedge}~
\llbracket A_p \rrbracket_{\overline{\phi}}~\tilde{\wedge}~\llbracket C \rrbracket_{\overline{\phi}} 
\leq \llbracket (u/x)D \rrbracket_{\overline{\phi}}$$
$$\llbracket A_1 \rrbracket_{\overline{\phi}}~\tilde{\wedge}~...~\tilde{\wedge}~
\llbracket A_p \rrbracket_{\overline{\phi}}~\tilde{\wedge}~\llbracket C \rrbracket_{\overline{\phi}} 
\leq \llbracket B \rrbracket_{\overline{\phi}}$$
Hence, as $\llbracket A_1 \rrbracket_{\overline{\phi}}~\tilde{\wedge}~...~\tilde{\wedge}~
\llbracket A_p \rrbracket_{\overline{\phi}} \leq \llbracket C \rrbracket_{\overline{\phi}}$, we have
$$\llbracket A_1 \rrbracket_{\overline{\phi}}~\tilde{\wedge}~...~\tilde{\wedge}~
\llbracket A_p \rrbracket_{\overline{\phi}} \leq \llbracket B \rrbracket_{\overline{\phi}}$$

\item If the last rule is {\bf Conversion}, then we use the fact that 
the model is a model of $\Sigma, {\cal R}$.
\end{itemize}
\end{proof}

\begin{corollary}
Let ${\cal M}$ be a model, valued in some $\Pi$-algebra ${\cal B}$,
of a theory $\Sigma, {\cal R}$.  
Then, for all judgments $\vdash t:B$ 
derivable in $\Sigma, {\cal R}$, we have
$\llbracket B \rrbracket_{\overline{\phi}} \geq \tilde{\top}$.
\end{corollary}

\medskip

\begin{corollary}
Let ${\cal M}$ be a model, valued in
the two-element $\Pi$-algebra of Example \ref{ex:01}, 
of a theory $\Sigma, {\cal R}$. 
Then, for all judgments $\vdash t:B$, derivable in this theory, 
we have $\llbracket B \rrbracket_{\overline{\phi}} = 1$.
\end{corollary}

\medskip

\begin{corollary}[Consistency]
Let $\Sigma, {\cal R}$ be a theory that has a model, valued in
the two-element $\Pi$-algebra of Example \ref{ex:01}.
Then, there is no term $t$ such that the judgment $P:Type \vdash t:P$
is derivable in $\Sigma, \Gamma$.
\end{corollary}

\section{Super-consistency}
\label{sec:sc}

\subsection{Super-consistency}

We now want to define a notion of notion of {\em super-consistency}: a
theory is super-consistent if for every $\Pi$-algebra, there exists a
model of this theory valued in this algebra.

Unfortunately, this constraint is sometimes too strong, as it does not
allow to define interpretations as fixed points, for instance if we
have a rule 
$$P \lra ((P \Rightarrow Q) \Rightarrow Q)$$
we want to
define the interpretation of $P$ as the fixed point of the function
mapping $b$ to $(b~\tilde{\Rightarrow}~a)~\tilde{\Rightarrow}~a$,
where $a$ is the interpretation of $Q$, but this function does not
have a fixed point in all $\Pi$-algebras. Thus, we weaken this
constraint, requiring the existence of model for {\em complete}
$\Pi$-algebras only. Defining this notion of completeness requires to
introduce an order relation $\sqsubseteq$, that need not be related to
the pre-order $\leq$.

\medskip

\begin{definition}[Ordered, complete $\Pi$-algebra]
A $\Pi$-algebra is {\em ordered} if it is equipped with an order relation 
$\sqsubseteq$ such that the operation $\tilde{\Pi}$ is left anti-monotonic
and right monotonic with respect to $\sqsubseteq$, that is
\begin{itemize}
\item if $a \sqsubseteq b$, then for all $S$, $\tilde{\Pi}(b,S) \sqsubseteq \tilde{\Pi}(a,S)$, 
\item if $S \sqsubseteq T$, then for all $a$, $\tilde{\Pi}(a,S) \sqsubseteq \tilde{\Pi}(a,T)$,
\end{itemize}
where the relation $\sqsubseteq$ is extended to sets of elements of 
${\cal B}$ in a trivial 
way: $S \sqsubseteq T$ if for all $a$ in $S$, there exists a $b$ in $T$ 
such that $a \sqsubseteq b$.

It is {\em complete} if every subset of ${\cal B}$ has a 
least upper bound for the relation $\sqsubseteq$.
\end{definition}

\begin{definition}[Super-consistency]
A theory $\Sigma, {\cal R}$, is {\em super-consistent} if,
for every full, ordered and complete $\Pi$-algebra ${\cal B}$, there
exists a model ${\cal M}$, valued in ${\cal B}$, of $\Sigma,
{\cal R}$.
\end{definition}

In the remainder of this section, we prove that the three theories
presented in Section \ref{sec:Examples} are super-consistent.

\subsection{Simple type theory}

Let ${\cal B} = \langle {\cal B}, \tilde{\top}, \tilde{\wedge}, 
{\cal P}^+({\cal B}), \tilde{\Pi} \rangle$ 
be a full $\Pi$-algebra. 
We construct a model of Simple type theory, valued in ${\cal B}$, in
two steps.  The first is the construction of the interpretation
function ${\cal M}$ and the proof of the validity of the congruence
for this function. The second is the construction of the
interpretation function $\llbracket . \rrbracket$ and the proof of the
validity of the congruence for this function.  The key idea in this
construction is to take ${\cal M}_o = {\cal B}$, to interpret
$\varepsilon$ as the identity over ${\cal B}$, and $\Rightarrow$ like
$\ra$ in order to validate the rewrite rule
$$\varepsilon~(\Rightarrow~x~y) \lra (\varepsilon~x) \ra (\varepsilon~y)$$

\begin{definition}
Let $S$ and $T$ be two sets, we write 
${\cal F}(S,T)$ for the set of functions from $S$ to $T$. 
\end{definition}

\medskip

\subsubsection{The interpretation function ${\cal M}$}

The first step of the proof is the construction of the interpretation 
function ${\cal M}$. 

Let $\{e\}$ be an arbitrary one-element set such that $e$ is not in ${\cal B}$.

\medskip

\begin{definition}
\label{MSTT}
The interpretation function ${\cal M}$ is defined as follows
\begin{itemize}
\item ${\cal M}_{Kind} = {\cal M}_{Type} = {\cal B}$,
\item ${\cal M}_{\Pi x:C~D} = {\cal F}({\cal M}_C, {\cal M}_D)$,
except if ${\cal M}_D = \{e\}$, in which case 
${\cal M}_{\Pi x:C~D} = \{e\}$,
\item ${\cal M}_{\iota} = {\cal M}_{\Rightarrow} = {\cal M}_{\fa_A} = 
{\cal M}_{\varepsilon} = \{e\}$, 
\item ${\cal M}_o = {\cal B}$,
\item ${\cal M}_x = \{e\}$, 
\item ${\cal M}_{\lambda x:C~t} = {\cal M}_t$,
\item ${\cal M}_{(t~u)} = {\cal M}_t$.
\end{itemize}
\end{definition}

\medskip

We first prove the two following lemmas. 

\begin{lemma}\label{lem1}
If the term $t$ is an object, then 
$${\cal M}_t = \{e\}$$
\end{lemma}

\begin{proof}
By induction on the structure of the term $t$. 
The term $t$ is neither $Kind$, $Type$, nor $o$. It is not a product. 
If it has the form $\lambda x:C~t'$, then $t'$ is an object.
If it has the form $(t'~t'')$, then $t'$ is an object.
\end{proof}

\begin{lemma}\label{lem2}
If $u$ is an object then
$${\cal M}_{(u/x)t} = {\cal M}_t$$
\end{lemma}

\begin{proof}
By induction on the structure of the term $t$. 
If $t = x$ then,
by Lemma \ref{lem1}
$${\cal M}_{(u/x)t} = {\cal M}_u = \{e\} = {\cal M}_t$$ If $t$ is
$Kind$, $Type$, a constant, or a variable different from $x$, then $x$
does not occur in $t$.  If it is a product, an abstraction, or an
application, we use the induction hypothesis.
\end{proof}

\begin{lemma}[Validity of the congruence]
If $t \equiv_{\beta {\cal R}} u$ then 
$${\cal M}_t = {\cal M}_u$$
\end{lemma}

\begin{proof}
If $t = ((\lambda x:C~t')~u')$, then $u'$ is an object and
by Lemma \ref{lem2}
$${\cal M}_{((\lambda x:C~t')~u')} = {\cal M}_{t'} = {\cal M}_{(u'/x)t'}$$
Then, as for all $v$, ${\cal M}_{(\varepsilon~v)} = {\cal M}_{\varepsilon} = \{e\}$, 
and if ${\cal M}_D = \{e\}$, then 
${\cal M}_{\Pi x:C~D} = \{e\}$, we have 
$${\cal M}_{(\varepsilon~C) \Rightarrow (\varepsilon~D)} = \{e\} = 
{\cal M}_{(\varepsilon~(C~\Rightarrow~D))}$$
and 
$${\cal M}_{\Pi x:C~(\varepsilon~(D~x))} = \{e\} = 
{\cal M}_{(\varepsilon~(\fa_C~D))}$$
We prove, by induction on $t$, that if 
$t \lra^1_{\beta {\cal R}} u$ then ${\cal M}_t = {\cal M}_u$
and we conclude with a simple induction on the structure of a reduction 
of $t$ and $u$ to a common term. 
\end{proof}

\subsubsection{The interpretation function $\llbracket . \rrbracket$}

The second step of the proof is the construction of the interpretation function
$\llbracket . \rrbracket$ and the proof of the validity of the
congruence for this function.

\medskip

\begin{definition}
\label{bracketSTT}
The interpretation function $\llbracket . \rrbracket$ is
defined as follows

\begin{itemize}
\item
$\llbracket Kind \rrbracket_{\phi} = \llbracket Type \rrbracket_{\phi} 
= \tilde{\top}$,

\item 
$\llbracket \Pi x:C~D \rrbracket_{\phi} = 
\tilde{\Pi}(\llbracket C \rrbracket_{\phi},
\{\llbracket D \rrbracket_{\phi, x = c}~|~c \in {\cal M}_C\})$,

\item 
$\llbracket \iota \rrbracket_{\phi} = \tilde{\top}$,

\item 
$\llbracket o \rrbracket_{\phi} = \tilde{\top}$,

\item $\llbracket \Rightarrow \rrbracket_{\phi}$ is 
the function mapping $a$ and $b$ in 
${\cal B}$ to $\tilde{\Pi}(a, \{b\})$, 

\item $\llbracket \fa_C \rrbracket_{\phi}$ is
the function mapping $f$ in ${\cal F}({\cal M}_C,{\cal B})$ to 
$\tilde{\Pi}(\llbracket C \rrbracket_{\phi}, \{(f~c)~|~c \in {\cal M}_C\})$, 

\item $\llbracket \varepsilon \rrbracket_{\phi}$ is the identity on
${\cal B}$,

\item $\llbracket x\rrbracket_{\phi} = \phi x$, 

\item $\llbracket \lambda x:C~t\rrbracket_{\phi}$ is the function
mapping $c$ in ${\cal M}_C$ to $\llbracket t\rrbracket_{\phi, x = c}$, except 
if for all $c$ in ${\cal M}_C$, $\llbracket t\rrbracket_{\phi, x = c} = e$ 
in which case $\llbracket \lambda x:C~t\rrbracket_{\phi} = e$, 

\item
$\llbracket (t~u)\rrbracket_{\phi} = 
\llbracket t \rrbracket_{\phi}~\llbracket u \rrbracket_{\phi}$, 
except if 
$\llbracket t \rrbracket_{\phi} = e$, in which case
$\llbracket (t~u)\rrbracket_{\phi} = e$.
\end{itemize}
\end{definition}

\medskip

\begin{lemma}[Well-typedness]
If $\Gamma \vdash t:B$, and $\phi$ is a function mapping variables
$x:A$ of $\Gamma$ to elements of ${\cal M}_A$, then
$$\llbracket t \rrbracket_{\phi} \in {\cal M}_B$$
\end{lemma}

\begin{proof}
We check each case of the definition of $\llbracket . \rrbracket$.
\end{proof}

\begin{lemma}[Substitution]
For all $t$, $u$ and $\phi$
$$\llbracket (u/x)t \rrbracket_{\phi} = \llbracket t \rrbracket_{\phi,
x = \llbracket u \rrbracket_{\phi}}$$
\end{lemma}

\begin{proof}
By induction on the structure of the term $t$. 
\end{proof}

\begin{lemma}[Validity of the congruence]
If $t \equiv_{\beta {\cal R}} u$ then 
$$\llbracket t \rrbracket_{\phi} = \llbracket u \rrbracket_{\phi}$$
\end{lemma}

\begin{proof}
If $t = ((\lambda x:C~t')~u')$, then 
if for all $c$ in ${\cal M}_C$, we have 
$\llbracket t' \rrbracket_{\phi, x = c} = e$, 
then
$$\llbracket ((\lambda x:C~t')~u') \rrbracket_{\phi} = e
= \llbracket t' \rrbracket_{\phi, x = \llbracket u' \rrbracket_{\phi}}
= \llbracket (u'/x)t' \rrbracket_{\phi}$$
Otherwise
$$\llbracket ((\lambda x:C~t')~u') \rrbracket_{\phi} = 
\llbracket t' \rrbracket_{\phi, x = \llbracket u' \rrbracket_{\phi}}
= \llbracket (u'/x)t' \rrbracket_{\phi}$$
Then 
$$\llbracket (\varepsilon~(\Rightarrow~t'~u')) \rrbracket_{\phi} = 
\tilde{\Pi}(
\llbracket t' \rrbracket_{\phi}, \{\llbracket u'
\rrbracket_{\phi} \})
= \llbracket (\varepsilon~t') \ra (\varepsilon~u') \rrbracket_{\phi}$$
and 
$$\llbracket (\varepsilon~(\fa_C~t')) \rrbracket_{\phi} = 
\tilde{\Pi}(\llbracket C \rrbracket_{\phi}, \{(\llbracket t' 
\rrbracket_{\phi}~c)~|~c \in {\cal M}_C\})
= \llbracket \Pi y:C~(\varepsilon~(t'~y)) \rrbracket_{\phi}$$
We prove, by induction on $t$, that if 
$t \lra^1_{\beta {\cal R}} u$ then 
$$\llbracket t \rrbracket_{\phi} = \llbracket u 
\rrbracket_{\phi}$$
and we conclude with a simple induction on the structure of a reduction 
of $t$ and $u$ to a common term. 
\end{proof}

We thus get the following theorem. 

\medskip

\begin{theorem}
Simple type theory is super-consistent.
\end{theorem}

\subsection{Simple type theory with a parametric quantifier}
\label{sttp}

In a model of Simple type theory with a parametric quantifier, like in
the previous section, we want to take ${\cal M}_o = {\cal
  B}$.  But, unlike in the previous section, we do not have $o:Type$,
but $o:type:Type$. So $o$ is now an object.

In the previous section, we took ${\cal M}_t = \{e\}$ for all objects.
This permitted to define ${\cal M}_{(t~u)}$ and ${\cal M}_{\lambda
x:C~t}$ as ${\cal M}_t$ and validate $\beta$-reduction
trivially. But this is not possible anymore in Simple type theory with
a parametric quantifier, where ${\cal M}_o$ is ${\cal B}$ and
${\cal M}_{arrow(o,o)}$ is ${\cal F}({\cal B},{\cal B})$. So,
we cannot define ${\cal M}_{\lambda x:type~x}$ to be 
${\cal M}_x$, but we need to define it 
as a function.
To help to construct this function, we need to construct first another
interpretation function $({\cal N}_t)_t$ and parametrize the definition of
${\cal M}_t$ itself by a function $\psi$ mapping variables of 
type $A$ to elements of ${\cal N}_A$. Thus the model is a three layer model.

Like in the previous section, we want to define ${\cal M}_{\Pi
  x:C~D, \psi}$, as the set of functions from ${\cal M}_{C, \psi}$ to
${\cal M}_{D,\psi'}$. But to define this set ${\cal M}_{D,\psi'}$, we need to extend
the function $\psi$, mapping $x$ to an element of ${\cal N}_C$.  To
have such an element of ${\cal N}_C$, we need to define ${\cal M}_{\Pi
  x:C~D, \psi}$ as the set of functions mapping $\langle c', c\rangle$
in ${\cal N}_C \times {\cal M}_{C, \psi}$ to an element of ${\cal
  M}_{D, (\psi, x = c')}$.  As a consequence, if $\phi$ is a function
mapping $x$ of type $A$ to some element of ${\cal M}_A$, we need to
define $\llbracket (t~u) \rrbracket_{\phi}$ not as $\llbracket t
\rrbracket_{\phi}~\llbracket u \rrbracket_{\phi}$ but as $\llbracket t
\rrbracket_{\phi}~\langle {\cal M}_{u, \psi}, \llbracket u
\rrbracket_{\phi}\rangle$.  As a consequence $\llbracket . \rrbracket$
must be parametrized by both $\psi$ and $\phi$.

Let ${\cal B} = \langle {\cal B}, \tilde{\top}, \tilde{\wedge}, 
{\cal P}^+({\cal B}), \tilde{\Pi} \rangle$ 
be a full $\Pi$-algebra. 

\subsubsection{The interpretation function ${\cal N}$}

The first step of the proof is the definition of the interpretation 
function 
${\cal N}$ and the proof of the validity of the congruence for this
function.

Let $\{e\}$ be an arbitrary one-element set.  Let ${\cal U}$ be a set
containing ${\cal B}$ and $\{e\}$, and closed by function space and 
Cartesian product, that is 
such that if $S$ and $T$ are in ${\cal U}$ then so are $S \times T$ and 
${\cal F}(S,T)$. 
Such a set can be constructed, with the replacement scheme, 
as follows
$${\cal U}_0 = \{{\cal B}, \{e\}\}$$
$${\cal U}_{n+1} = {\cal U}_n 
\cup \{S \times T~|~S,T \in {\cal U}_n\}
\cup \{{\cal F}(S,T)~|~S,T \in {\cal U}_n\}$$
$${\cal U} = \bigcup_n {\cal U}_n$$

Then, let ${\cal V}$ be the smallest set containing $\{e\}$, ${\cal
B}$, and ${\cal U}$, and closed by Cartesian product and dependent
function space, that is, if $S$ is in ${\cal V}$ and $T$ is a family
of elements of ${\cal V}$ indexed by $S$, then the set of functions
mapping an element $s$ of $S$ to an element of $T_s$ is an element of
${\cal V}$.  As noted in \cite{MiquelWerner}, the construction of the
set ${\cal V}$, unlike that of ${\cal U}$, requires an inaccessible
cardinal.
Note that ${\cal U}$ is both an element and a subset of ${\cal V}$. 

\medskip

\begin{definition}
The interpretation function ${\cal N}$ is defined as follows

\begin{itemize}
\item ${\cal N}_{Type} = {\cal N}_{Kind} = {\cal V}$,
\item ${\cal N}_{\Pi x:C~D}$ is the set ${\cal F}({\cal N}_C,{\cal N}_{D})$, 
except if ${\cal N}_{D} = \{e\}$, in which case ${\cal N}_{\Pi x:C~D} = \{e\}$,
\item ${\cal N}_{type} = {\cal U}$,
\item ${\cal N}_{\iota} = {\cal N}_{o} = {\cal N}_{arrow} = 
{\cal N}_{\Rightarrow} = {\cal N}_{\fa} = {\cal N}_{\eta} = 
{\cal N}_{\varepsilon} = \{e\}$, 
\item ${\cal N}_{x} = \{e\}$, 
\item ${\cal N}_{\lambda x:C~t} = {\cal N}_{t}$,
\item ${\cal N}_{(t~u)} = {\cal N}_{t}$.
\end{itemize}
\end{definition}

\medskip

We first prove the two following lemmas.

\begin{lemma}\label{lem1sttp}
If the term $t$ is an object, then 
$${\cal N}_t = \{e\}$$
\end{lemma}

\begin{proof}
By induction on the structure of the term $t$. 
The term $t$ is neither $Kind$, $Type$, nor $type$. It is not a product. 
If it has the form $\lambda x:C~t'$, then $t'$ is an object.
If it has the form $(t'~t'')$, then $t'$ is an object.
\end{proof}

\begin{lemma}\label{lem2sttp}
If $u$ is an object, then 
$${\cal N}_{(u/x)t} = {\cal N}_t$$
\end{lemma}

\begin{proof}
By induction on the structure of the term $t$. 
If $t = x$ then,
by Lemma \ref{lem1sttp}
$${\cal N}_{(u/x)t} = {\cal N}_u = \{e\} = {\cal N}_t$$
If $t$ is $Kind$, $Type$, a constant, 
or a variable different from $x$, then $x$ does not occur in $t$. 
If it is a product, an abstraction, or an application, 
we use the induction hypothesis.
\end{proof}

\begin{lemma}[Validity of the congruence]
\label{prop:validity1}
If $t \equiv_{\beta {\cal R}} u$ then
$${\cal N}_t = {\cal N}_u$$
\end{lemma}

\begin{proof}
If $t = ((\lambda x:C~t')~u')$, then $u'$ is an object and by 
Lemma \ref{lem2sttp}
$${\cal N}_{((\lambda x:C~t')~u')} = {\cal N}_{t'} = {\cal N}_{(u'/x)t'}$$
Then, as for all $v$, ${\cal N}_{(\eta~v)} = {\cal N}_{\eta} = \{e\}$ and if 
${\cal N}_{D} = \{e\}$, then ${\cal N}_{\Pi x:C~D} = \{e\}$, we have 
$${\cal N}_{(\eta~(arrow~C~D))} = \{e\} = {\cal N}_{((\eta~C) \ra (\eta~D))}$$
As for all $v$, ${\cal N}_{(\varepsilon~v)} = {\cal N}_{\varepsilon} = \{e\}$, 
and if ${\cal N}_{D} = \{e\}$, then ${\cal N}_{\Pi x:C~D} = \{e\}$, we have 
$${\cal N}_{(\varepsilon~(\Rightarrow~C~D))} = \{e\} = 
{\cal N}_{((\varepsilon~C) \ra (\varepsilon~D))}$$
and 
$${\cal N}_{(\varepsilon~(\fa~C~D))} = \{e\} = 
{\cal N}_{\Pi x:(\eta~C)~(\varepsilon~(D~x))}$$
We prove, by induction on $t$, that if $t \lra^1_{\beta {\cal R}} u$
then ${\cal N}_{t} = {\cal N}_{u}$ 
and we conclude with a simple induction on the structure of a reduction 
of $t$ and $u$ to a common term. 
\end{proof}

\subsubsection{The interpretation function ${\cal M}$}

The second step of the proof is the definition of the 
interpretation function 
${\cal M}$ and the proof of the validity of the congruence for this
function.

\medskip

\begin{definition}

The interpretation function ${\cal M}$ is defined as follows

\begin{itemize}
\item 
${\cal M}_{Kind,\psi} = {\cal M}_{Type,\psi} = {\cal B}$,

\item
${\cal M}_{\Pi x:C~D, \psi, \phi}$ is the set of functions $f$ mapping 
$\langle c', c \rangle$ in 
${\cal N}_C \times {\cal M}_{C, \psi}$ to an element of 
${\cal M}_{D, (\psi, x = c')}$, except if for all $c'$ in ${\cal N}_C$, 
${\cal M}_{D, (\psi, x = c')} = \{e\}$, in which case 
${\cal M}_{\Pi x:C~D, \psi} = \{e\}$, 

\item ${\cal M}_{type,\psi} = {\cal B}$,

\item ${\cal M}_{\eta,\psi}$ is the function of ${\cal F}({\cal U},{\cal V})$ 
mapping $S$ to $S$,

\item ${\cal M}_{\varepsilon,\psi}$ is the function of
${\cal F}(\{e\}, {\cal V})$, mapping $e$ to $\{e\}$, 

\item ${\cal M}_{\iota,\psi} = \{e\}$,

\item ${\cal M}_{o,\psi} = {\cal B}$,

\item ${\cal M}_{arrow, \psi}$ is the function mapping $S$ and $T$
in ${\cal U}$ to the set ${\cal F}(\{e\} \times S,T)$,
except if $T = \{e\}$ in which case it maps $S$ and $T$ to $\{e\}$, 

\item ${\cal M}_{\Rightarrow, \psi} = {\cal M}_{\fa, \psi} = e$, 

\item ${\cal M}_{x,\psi} = \psi x$, 

\item ${\cal M}_{\lambda x:C~t,\psi}$ is the function mapping $c$ in 
${\cal N}_C$ to ${\cal M}_{t, (\psi, x = c)}$, except if for all $c$ in 
${\cal N}_C$, ${\cal M}_{t, (\psi,  x = c)} = e$, in which case
${\cal M}_{\lambda x:C~t,\psi} = e$,

\item ${\cal M}_{(t~u),\psi} = {\cal M}_{t,\psi}~{\cal M}_{u,\psi}$,
except if ${\cal M}_{t,\psi} = e$ in which case ${\cal M}_{(t~u),\psi} = e$. 
\end{itemize}
\end{definition}

\medskip

\begin{lemma}\label{limitedsttp}
If $\Gamma \vdash C:Type$, then 
$${\cal N}_C \in {\cal V}$$
\end{lemma}

\begin{proof}
By induction on the structure of the term $C$. As this term has type
$Type$, it is neither $Kind$ nor $Type$.
\end{proof}

\begin{lemma}[Well-typedness]
\label{wt1sttp}
If $\Gamma \vdash t:B$ and $\psi$ is a function mapping the variables
$x:A$ of $\Gamma$ to elements of ${\cal N}_A$, then 
$${\cal M}_{t,\psi} \in {\cal N}_B$$
\end{lemma}

\begin{proof}
We check each case of the definition of ${\cal M}$.
\end{proof}

\begin{lemma}[Substitution]
\label{subst1sttp}
For all $t$, $u$ and $\psi$
$${\cal M}_{(u/x)t,\psi} = {\cal M}_{t, (\psi, x = {\cal M}_u)}$$
\end{lemma}

\begin{proof}
By induction on the structure of the term $t$. 
\end{proof}

\begin{lemma}[Validity of the congruence]
\label{prop:validity2}
If $t \equiv_{\beta {\cal R}} u$ then 
$${\cal M}_{t,\psi} = {\cal M}_{u,\psi}$$
\end{lemma}

\begin{proof}
If $t = ((\lambda x:C~t')~u')$, then 
if 
for all $c$ in ${\cal N}_C$ 
${\cal M}_{t', (\psi , x = c)} = e$, then 
$${\cal M}_{((\lambda x:C~t')~u'),\psi} = e = 
{\cal M}_{t', (\psi, x = {\cal M}_{u',\psi})} = {\cal M}_{(u'/x)t',\psi}$$
Otherwise 
$${\cal M}_{((\lambda x:C~t')~u'),\psi} 
= {\cal M}_{t', (\psi, x = {\cal M}_{u',\psi})} = {\cal M}_{(u'/x)t',\psi}$$
The set ${\cal M}_{(\eta~(arrow~C~D)),\psi}$
is the set ${\cal F}(\{e\} \times {\cal M}_{C,\psi},{\cal M}_{D,\psi})$, 
except if ${\cal M}_{D,\psi} = \{e\}$, in which case 
${\cal M}_{(\eta~(arrow~C~D)),\psi} = \{e\}$.
The set ${\cal M}_{((\eta~C) \ra (\eta~D)),\psi}$
is this same set.
Thus 
$${\cal M}_{(\eta~(arrow~C~D)),\psi} = {\cal M}_{((\eta~C) \ra (\eta~D)),\psi}$$
We have 
$${\cal M}_{(\varepsilon~(\Rightarrow~C~D)), \psi} = \{e\} = 
{\cal M}_{((\varepsilon~C) \ra (\varepsilon~D)), \psi}$$
and
$${\cal M}_{(\varepsilon~(\fa~C~D)), \psi} = \{e\} = 
{\cal M}_{\Pi x:(\varepsilon~C)~(\varepsilon~(D~x)), \psi}$$
We prove, by induction on $t$, that if 
$t \lra^1_{\beta {\cal R}} u$ then ${\cal M}_{t,\psi} = {\cal M}_{u,\psi}$
and we conclude with a simple induction on the structure of a reduction 
of $t$ and $u$ to a common term. 
\end{proof}

\subsubsection{The interpretation function $\llbracket . \rrbracket$}

The last step of the proof is the definition of the interpretation
function $\llbracket . \rrbracket$ and the proof of the validity of
the congruence for this function.

\medskip

\begin{definition}
The interpretation function $\llbracket . \rrbracket$ is defined as
follows

\begin{itemize}
\item
$\llbracket Kind\rrbracket_{\psi,\phi} = \llbracket Type\rrbracket_{\psi,\phi} =
\tilde{\top}$,

\item 
$\llbracket \Pi x:C~D \rrbracket_{\psi,\phi} = 
\tilde{\Pi}(\llbracket C \rrbracket_{\psi,\phi},
\{\llbracket D \rrbracket_{
(\psi, x = c'),(\phi, x = c)}~|~c' \in {\cal N}_C, 
c \in {\cal M}_{C,\psi}\})$,

\item 
$\llbracket type \rrbracket_{\psi,\phi} = \tilde{\top}$,

\item 
$\llbracket \iota \rrbracket_{\psi,\phi} = \tilde{\top}$,

\item 
$\llbracket o \rrbracket_{\psi,\phi} = \tilde{\top}$, 

\item 
$\llbracket arrow \rrbracket_{\psi,\phi}$ 
is the function from ${\cal U} \times {\cal B}$ and ${\cal U} \times {\cal B}$ 
to ${\cal B}$ mapping $\langle S, a \rangle$ 
and $\langle T, b \rangle$ to $\tilde{\Pi}(a, \{b\})$, 

\item 
$\llbracket \Rightarrow \rrbracket_{\psi,\phi}$ is the function $\{e\}
\times {\cal B}$ and $\{e\} \times {\cal B}$ to ${\cal B}$ mapping
$\langle e, a \rangle$ and $\langle e, b \rangle$ to $\tilde{\Pi}(a,
\{b\})$,

\item $\llbracket \fa \rrbracket_{\psi,\phi}$ is the
function 
mapping
$\langle S, a \rangle$ in ${\cal U} \times {\cal B}$, 
and $\langle e, g \rangle$ in $\{e\} \times 
{\cal F}(\{e\} \times S, {\cal B})$ to $\tilde{\Pi}(a, 
\{(g~\langle e,s \rangle)~|~s \in S\})$,

\item $\llbracket \eta \rrbracket_{\psi,\phi}$ is 
the function from ${\cal U} \times 
{\cal B}$ to ${\cal B}$, mapping $\langle S, a \rangle$ to $a$, 

\item $\llbracket \varepsilon \rrbracket_{\psi,\phi}$
is the function from 
$\{e\} \times {\cal B}$ to ${\cal B}$,
mapping $\langle e, a \rangle$ to $a$, 

\item $\llbracket x\rrbracket_{\psi,\phi} = \phi x$, 

\item
$\llbracket \lambda x:C~t\rrbracket_{\psi,\phi}$ is the function
mapping 
$\langle c', c \rangle$ in ${\cal N}_C \times {\cal M}_{C, \psi}$ 
to $\llbracket t\rrbracket_{(\psi,x=c'),(\phi, x = c)}$,
except if 
for all $\langle c', c \rangle$ in ${\cal N}_C \times {\cal M}_{C, \psi}$, 
$\llbracket t \rrbracket_{(\psi, x = c'),(\phi, x = c)} = e$, 
in which case $\llbracket \lambda x:C~t\rrbracket_{\psi,\phi} = e$, 

\item
$\llbracket (t~u)\rrbracket_{\psi,\phi} = 
\llbracket t\rrbracket_{\psi,\phi}~
\langle {\cal M}_{u, \psi}, \llbracket u \rrbracket_{\psi,\phi}\rangle$, 
except if $\llbracket t \rrbracket_{\psi,\phi} = e$, in which case
$\llbracket (t~u)\rrbracket_{\psi,\phi} = e$.
\end{itemize}
\end{definition}

\medskip

\begin{lemma}[Well-typedness]
\label{wt2sttp}
If $\Gamma \vdash t:B$, 
$\psi$ is a function mapping variables $x:A$ of $\Gamma$ to 
elements of ${\cal N}_A$, and $\phi$ is a function mapping variables 
$x:A$ of $\Gamma$ to elements of ${\cal M}_{A,\psi}$, 
then 
$$\llbracket t \rrbracket_{\psi,\phi} \in {\cal M}_{B,\psi}$$
\end{lemma}

\begin{proof}
We check each case of the definition of $\llbracket . \rrbracket$.

Let us check, for instance, that if $t = (t_1~t_2)$, 
$t_1$ has type $\Pi x:C~D$ and 
$t_2$ has type $C$, hence $(t_1~t_2)$ has type $(t_2/x)D$, 
then 
$\llbracket (t_1~t_2) \rrbracket_{\psi,\phi}$ is in
${\cal M}_{(t_2/x)D, \psi}$. 
We have ${\cal M}_{t_2, \psi}$ is in ${\cal N}_C$ and, 
by induction hypothesis, 
$\llbracket t_1\rrbracket_{\psi,\phi}$ is in ${\cal M}_{\Pi x:C~D, \psi}$, 
and $\llbracket t_2 \rrbracket_{\psi,\phi}$ is in ${\cal M}_{C, \psi}$. 
We have $\llbracket (t_1~t_2) \rrbracket_{\psi,\phi} = 
\llbracket t_1 \rrbracket_{\psi,\phi}~\langle {\cal M}_{t_2, \psi}, 
\llbracket t_2 \rrbracket_{\psi,\phi} \rangle$
and, by definition of ${\cal M}_{\Pi x:C~D, \psi}$, this term
is in ${\cal M}_{D, (\psi, x = {\cal M}_{t_2, \psi})}$, that is 
${\cal M}_{(t_2/x)D, \psi}$.
\end{proof}

\begin{lemma}[Substitution]
\label{subst2sttp}
For all $t$, $u$, $\psi$, and $\phi$
$$\llbracket (u/x)t \rrbracket_{\psi, \phi} = \llbracket t \rrbracket_{
\psi, x = {\cal M}_{u,\psi}
\phi, x = \llbracket u \rrbracket_{\psi,\phi}}$$
\end{lemma}

\begin{proof}
By induction on the structure of the term $t$.
\end{proof}

\begin{lemma}[Validity of the congruence]
\label{prop:validity3}
If $t \equiv_{\beta {\cal R}} u$ then 
$$\llbracket t \rrbracket_{\psi,\phi} = \llbracket u \rrbracket_{\psi,\phi}$$
\end{lemma}

\begin{proof}
If $t = ((\lambda x:C~t')~u')$, then 
if 
for all $c'$ in ${\cal N}_C$ and 
$c$ in ${\cal M}_{C, \psi}$ we have 
$\llbracket t' \rrbracket_{(\psi,x=c'),(\phi, x = c)} = e$
then
$$\llbracket ((\lambda x:C~t')~u') \rrbracket_{\psi,\phi} = 
e = \llbracket t' \rrbracket_{(\psi, x = {\cal M}_{u',\psi}),(\phi, x =
\llbracket u'\rrbracket_{\psi,\phi})}
= \llbracket (u'/x)t' \rrbracket_{\psi,\phi}$$
Otherwise
$$\llbracket ((\lambda x:C~t')~u') \rrbracket_{\psi,\phi} = 
\llbracket t' \rrbracket_{(\psi, x = {\cal M}_{u', \psi}),(\phi, x = 
\llbracket u' \rrbracket_{\psi,\phi})} 
= \llbracket (u'/x)t' \rrbracket_{\psi,\phi}$$
We have 
$$\llbracket (\eta~(arrow~C~D)) \rrbracket_{\psi,\phi} = 
\tilde{\Pi}(\llbracket C \rrbracket_{\psi,\phi}, \{\llbracket D \rrbracket_{\psi,\phi} \})
= \llbracket (\eta~C) \ra (\eta~D) \rrbracket_{\psi,\phi}$$
$$\llbracket (\varepsilon~(\Rightarrow~C~D)) \rrbracket_{\psi,\phi} = 
\tilde{\Pi}(\llbracket C \rrbracket_{\psi,\phi}, \{\llbracket D \rrbracket_{\psi,\phi} 
\}) = \llbracket (\varepsilon~C) \ra (\varepsilon~D) \rrbracket_{\psi,\phi}$$
$$\llbracket (\varepsilon~(\fa~C~D)) \rrbracket_{\psi,\phi} = 
\tilde{\Pi}(\llbracket C \rrbracket_{\psi,\phi}, \{(\llbracket D \rrbracket_{\psi,\phi}~\langle e, c \rangle)~|~c \in {\cal M}_{C,\psi}\})
= \llbracket \Pi y:(\eta~C)~(\varepsilon~(D~y)) \rrbracket_{\psi,\phi}$$
We prove, by induction on $t$, that if 
$t \lra^1_{\beta {\cal R}} u$ then $\llbracket t \rrbracket_{\psi,\phi} = \llbracket u 
\rrbracket_{\psi,\phi}$
and we conclude with a simple induction on the structure of a reduction 
of $t$ and $u$ to a common term. 
\end{proof}

We thus get the following theorem. 

\medskip

\begin{theorem}
Simple type theory with a parametric quantifier is super-consistent.
\end{theorem}

\medskip

\begin{remark}
The set ${\cal V}$, thus an inaccessible cardinal, are not
really needed to prove the super-consistency of
Simple type theory with a parametric quantifier if we can 
adapt the notion of model in such a way that the family
${\cal N}$ is defined for type families only. 
Then, Lemma
\ref{wt1sttp} is proved for objects only. This is sufficient to 
define ${\cal M}_{\lambda x:type~x, \psi}$ as the identity on ${\cal U}$ and, 
more generally, the function ${\cal M}$. 
In this case, the class of sets ${\cal M}_{t,\psi}$ would not be a set, 
which is common in models of many sorted Predicate logic 
with an infinite number of sorts.

The systematic development of this notion of 
partial interpretation is left for future work.
\end{remark}

\subsection{The Calculus of constructions}

A very similar proof can be made for the Calculus of constructions.

In the construction of the interpretation functions
${\cal N}$, ${\cal M}$, and $\llbracket . \rrbracket$, we drop 
the clauses associated to the symbols $\iota$, $\Rightarrow$, $\fa$
and $arrow$ and we add the clauses. 

\begin{itemize}
\item ${\cal N}_{\dot{\Pi}_{TT}} = 
{\cal N}_{\dot{\Pi}_{TK}} = {\cal N}_{\dot{\Pi}_{KT}} = 
{\cal N}_{\dot{\Pi}_{KK}}  = \{e\}$

\item ${\cal M}_{\dot{\Pi}_{KK}}$ is the function 
mapping $S$ in ${\cal U}$ and $h$ in ${\cal F}(\{e\}, {\cal U})$ 
to the set ${\cal F}(\{e\} \times S,(h~e))$, 
except if $(h~e) = \{e\}$ in which case it maps $S$ and $h$ to $\{e\}$, 

\item ${\cal M}_{\dot{\Pi}_{TT}} = e$, 

\item ${\cal M}_{\dot{\Pi}_{KT}} = e$, 

\item ${\cal M}_{\dot{\Pi}_{TK}}$ is the function 
mapping $e$ and $h$ in ${\cal F}(\{e\}, {\cal U})$ 
to the set ${\cal F}(\{e\}\times\{e\}, (h~e))$, except if $(h~e) = \{e\}$ 
in which case it maps $e$ and $h$ to $\{e\}$, 

\item $\llbracket \dot{\Pi}_{KK} \rrbracket_{\psi,\phi}$ 
is the function mapping 
$\langle S, a \rangle$ in ${\cal U} \times {\cal B}$, 
$\langle f,g\rangle$ in 
${\cal F}(\{e\}, {\cal U}) \times {\cal F}(\{e\} \times S, {\cal B})$ to
$\tilde{\Pi}(a,\{(g~\langle e, s\rangle)~|~s \in S\})$,

\item $\llbracket \dot{\Pi}_{TT} \rrbracket_{\psi,\phi}$ 
is the function mapping 
$\langle e, a \rangle$ in $\{e\} \times {\cal B}$, 
and $\langle e, g\rangle$ in $\{e\} \times {\cal F}(\{e\} \times \{e\}, 
{\cal B})$ to $\tilde{\Pi}(a,\{(g~\langle e, e \rangle)\})$,

\item $\llbracket \dot{\Pi}_{KT} \rrbracket_{\psi,\phi}$ 
is the function mapping 
$\langle S,a \rangle$ in ${\cal U} \times {\cal B}$, 
and $\langle e, g \rangle$ in $\{e\} \times {\cal F}(\{e\} \times S, {\cal B})$ 
to $\tilde{\Pi}(a,\{(g~\langle e, s \rangle)~|~s \in S\})$,

\item $\llbracket \dot{\Pi}_{TK} \rrbracket_{\psi,\phi}$
is the function mapping 
$\langle e, a \rangle$ in $\{e\} \times {\cal B}$, 
and $\langle f, g \rangle$ in ${\cal F}(\{e\},{\cal U}) \times
{\cal F}(\{e\} \times \{e\}, {\cal B})$ to
$\tilde{\Pi}(a,\{(g~\langle e, e \rangle)\})$.
\end{itemize}

The proof of Lemmas \ref{lem1sttp} and \ref{lem2sttp} are similar. 

The proof of Lemma \ref{prop:validity1} is similar, except 
for the case of rewrite rules.
$${\cal N}_{(\eta~(\dot{\Pi}_{KK}~C~D))} = \{e\} 
= {\cal N}_{\Pi x:(\eta~C)~(\eta~(D~x))}$$
$${\cal N}_{(\varepsilon~(\dot{\Pi}_{TT}~C~D))} = \{e\} 
= {\cal N}_{\Pi x:(\varepsilon~C)~(\varepsilon~(D~x))}$$
$${\cal N}_{(\varepsilon~(\dot{\Pi}_{KT}~C~D))} = \{e\} 
= {\cal N}_{\Pi x:(\eta~C)~(\varepsilon~(D~x))}$$
$${\cal N}_{(\eta~(\dot{\Pi}_{TK}~C~D))} = \{e\} 
= {\cal N}_{\Pi x:(\varepsilon~C)~(\eta~(D~x))}$$

The proof of Lemma \ref{limitedsttp} is similar.

The proof of Lemma \ref{wt1sttp} must be adapted to 
check the case of the symbols 
$\dot{\Pi}_{KK}$, 
$\dot{\Pi}_{TT}$, 
$\dot{\Pi}_{KT}$, and
$\dot{\Pi}_{TK}$.

The proof of Lemma \ref{subst1sttp} is similar.

The proof of Lemma \ref{prop:validity2} is similar, except 
for the case of rewrite rules.

The set 
${\cal M}_{(\eta~(\dot{\Pi}_{KK}~C~D)),\psi}$ 
is the set ${\cal F}(\{e\} \times {\cal M}_{C,\psi},({\cal M}_{D,\psi}~e))$, except if 
$({\cal M}_{D,\psi}~e) = \{e\}$ in which case 
${\cal M}_{(\eta~(\dot{\Pi}_{KK}~C~D)),\psi} = 
\{e\}$. The set ${\cal M}_{\Pi x:(\eta~C)~(\eta~(D~x)),\psi}$
is this same set. Thus
$${\cal M}_{(\eta~(\dot{\Pi}_{KK}~C~D)),\psi} 
= {\cal M}_{\Pi x:(\eta~C)~(\eta~(D~x)),\psi}$$
We have 
$${\cal M}_{(\varepsilon~(\dot{\Pi}_{TT}~C~D))} = \{e\}
= {\cal M}_{\Pi x:(\varepsilon~C)~(\varepsilon~(D~x))}$$
We have 
$${\cal M}_{(\varepsilon~(\dot{\Pi}_{KT}~C~D))} = \{e\} 
= {\cal M}_{\Pi x:(\varepsilon~C)~(\varepsilon~(D~x))}$$
The set 
${\cal M}_{(\eta~(\dot{\Pi}_{TK}~C~D)),\psi}$ 
is the set ${\cal F}(\{e\} \times \{e\},({\cal M}_{D,\psi}~e))$, except if 
$({\cal M}_{D,\psi}~e) = \{e\}$ in which case 
${\cal M}_{(\eta~(\dot{\Pi}_{TK}~C~D)),\psi} = 
\{e\}$. The set ${\cal M}_{\Pi x:(\varepsilon~C)~(\eta~(D~x)),\psi}$
is this same set. Thus
$${\cal M}_{(\eta~(\dot{\Pi}_{TK}~C~D)),\psi} 
= {\cal M}_{\Pi x:(\varepsilon~C)~(\eta~(D~x)),\psi}$$

The proof of Lemma \ref{wt2sttp}
must be adapted to 
check the case of the symbols 
$\dot{\Pi}_{KK}$, 
$\dot{\Pi}_{TT}$, 
$\dot{\Pi}_{KT}$, and
$\dot{\Pi}_{TK}$.

The proof of Lemma \ref{subst2sttp} is similar.

The proof of Lemma \ref{prop:validity3} is similar, except 
for the case of rewrite rules.
$$\llbracket (\eta~(\dot{\Pi}_{KK}~C~D)) \rrbracket_{\psi,\phi}
= \tilde{\Pi}(\llbracket C \rrbracket_{\psi,\phi}, \{(\llbracket D \rrbracket_{\psi,\phi}~\langle e, s \rangle)~|~s \in {\cal M}_{C,\psi}\})
= \llbracket \Pi y:(\eta~C)~(\eta~(D~y)) \rrbracket_{\psi,\phi}$$
$$\llbracket (\varepsilon~(\dot{\Pi}_{TT}~C~D)) \rrbracket_{\psi,\phi} = 
\tilde{\Pi}(\llbracket C \rrbracket_{\psi,\phi}, \{(\llbracket D \rrbracket_{\psi,\phi}~\langle e, e \rangle)\})
= \llbracket \Pi y:(\varepsilon~C)~(\varepsilon~(D~y)) \rrbracket_{\psi,\phi}$$
$$\llbracket (\varepsilon~(\dot{\Pi}_{KT}~C~D)) \rrbracket_{\psi,\phi}
= \tilde{\Pi}(\llbracket C \rrbracket_{\psi,\phi}, \{(\llbracket D \rrbracket_{\psi,\phi}~\langle e, s \rangle)~|~s \in {\cal M}_{C,\psi}\})
= \llbracket \Pi y:(\eta~C)~(\varepsilon~(D~y)) \rrbracket_{\psi,\phi}$$
$$\llbracket (\eta~(\dot{\Pi}_{TK}~C~D)) \rrbracket_{\psi,\phi}
= \tilde{\Pi}(\llbracket C \rrbracket_{\psi,\phi}, 
\{(\llbracket D \rrbracket_{\psi,\phi}\langle e, e \rangle)\})
= \llbracket \Pi y:(\varepsilon~C)~(\eta~(D~y)) \rrbracket_{\psi,\phi}$$

\section{Termination of proof reduction}
\label{sec:term}

We finally prove that proof reduction terminates in the $\lambda \Pi$-calculus
modulo any super-consistent theory such as Simple type theory 
without or with a parametric quantifier or the 
Calculus of constructions. 

In Deduction modulo theory, we can define a congruence with non terminating
rewrite rules, without affecting the
termination of proof reduction. For instance, the rewrite rule 
$$c \lra c$$ 
does not terminate, but the congruence it defines 
is the identity and proofs modulo this congruence are just proofs in
pure Predicate logic.  Thus, proof reduction in Deduction modulo this
congruence terminates. 
So, in the $\lambda \Pi$-calculus modulo this congruence, 
the $\beta$-reduction
terminates, but the $\beta{\cal R}$-reduction does not, as
the ${\cal R}$-reduction alone does not terminate.
Here, we restrict to prove the termination of
$\beta$-reduction, not $\beta {\cal R}$-reduction. In some cases,
like for the three theories presented above, the
termination of the $\beta {\cal R}$-reduction is a simple corollary of
the termination of the $\beta$-reduction.
In some others, it is not.

The main notion used in this proof is that of {\em reducibility 
candidate} introduced by Girard \cite{Girard}. Our inductive definition, 
however, follows that of Parigot \cite{Parigot}.

\subsection{The candidates}

\begin{definition}[Operations on set of terms]
The set $\tilde{\top}$ is defined as the set of strongly terminating terms.

Let $C$ be a set of terms and $S$ be a set of sets of terms. 
The set $\tilde{\Pi}(C,S)$ is defined as the set of strongly terminating 
terms $t$ such that if $t \lra_{\beta}^* \lambda x:A~t'$ then for all 
$t''$ in $C$, 
and for all $D$ in $S$, $(t''/x)t' \in D$. 
\end{definition}

The main property of the operation $\tilde{\Pi}$ is expressed by the following 
Lemma.

\begin{lemma}\label{application0}
Let $C$ be a set of terms and $S$ be a set of sets of terms, 
$t_1$, $t_2$, and $u$ be terms such that 
$t_1 \in \tilde{\Pi}(C,S)$,
$t_2 \in C$, 
and $(t_1~t_2) \lra_{\beta}^1 u$, 
$n_1$ and $n_2$ be natural numbers such that 
$n_1$ is the maximum length of
a reduction sequence issued from $t_1$, and $n_2$ is the maximum length
of a reduction sequence issued from $t_2$, and $D$ be an element of $S$.
Then, $u \in D$.
\end{lemma}

\begin{proof}
By induction on $n_1 + n_2$. 
If the reduction is at the root of the term, then 
$t_1$ has the form $\lambda x:A~t'$ and 
$u = (t_2/x)t'$.  By the definition of $\tilde{\Pi}(C,S)$, $u \in D$.
Otherwise, the reduction takes place in $t_1$ or in $t_2$, and
we apply the induction hypothesis.
\end{proof}

\begin{definition}[Candidates]
{\em Candidates} are inductively defined by the three rules 
\begin{itemize}
\item the set $\tilde{\top}$ of all strongly terminating terms is a candidate,
\item if $C$ is a candidate and $S$ is a set of 
candidates, then $\tilde{\Pi}(C,S)$ is a candidate,
\item if $S$ is a non empty set of candidates, then $\bigcap S$ 
is a candidate.
\end{itemize}
We write ${\cal C}$ for the set of candidates. 
\end{definition}

The algebra
$\langle {\cal C}, \leq, \tilde{\top}, \tilde{\wedge}, {\cal
  P}^{+}({\cal C}), \tilde{\Pi} \rangle$,
where $\leq$ is the trivial relation such that $C \leq C'$ always, and
$\tilde{\wedge}$ is any function from ${\cal C} \times {\cal C}$ to
${\cal C}$, for instance the constant function equal to
$\tilde{\top}$, is a full $\Pi$-algebra.

It is ordered by the subset relation and complete for this order.

\medskip

\begin{lemma}[Termination]
If $C$ is a candidate, then all the elements of $C$ strongly terminate.
\end{lemma}

\begin{proof}
By induction on the construction of $C$.
\end{proof}

\begin{lemma}[Variables]\label{Varpi}
If $C$ is a candidate and $x$ is a variable, 
then $x \in C$.
\end{lemma}

\begin{proof}
By induction on the construction of $C$.
\end{proof}

\begin{lemma}[Closure by reduction]\label{closurepi}
If $C$ is a candidate, $t \in C$, and $t \lra_{\beta}^* t'$, then $t' \in C$.
\end{lemma}

\begin{proof}
By induction on the construction of $C$.

If $C = \tilde{\top}$, then as $t$ is an element of $C$, it strongly terminates, 
thus $t'$ strongly terminates, and $t' \in C$. 

If $C = \tilde{\Pi}(D,S)$, 
then as $t$ is an element of $C$, it strongly terminates, 
thus $t'$ strongly terminates. 
If moreover $t' \lra_{\beta}^* \lambda x:A~t_1$, then $t \lra_{\beta}^* \lambda x:A~t_1$, 
and for all $u$ in $D$, 
and for all ${\cal U}$ in $S$, $(u/x)t_1 \in {\cal U}$. 
Thus, $t' \in C$.

If $C = \bigcap_i C_i$, 
then for all $i$, $t \in C_i$ and by induction hypothesis $t' \in C_i$.
Thus, $t' \in C$.
\end{proof}

\begin{lemma}[Applications]\label{application}
Let $C$ be a candidate and $S$ be a set of candidates, $t_1$ and
$t_2$ such that $t_1 \in \tilde{\Pi}(C,S)$ and $t_2 \in C$, and $D$
be an element of $S$. Then $(t_1~t_2) \in D$.
\end{lemma}

\begin{proof}
As $t_1 \in \tilde{\Pi}(C,S)$ and and $t_2 \in C$, $t_1$ and
$t_2$ strongly terminate.  Let $n_1$ be the maximum length of
a reduction sequence issued from $t_1$ and $n_2$ be the maximum
length of a reduction sequence issued from $t_2$. By Lemma
\ref{application0}, all the one step reducts of $(t_1~t_2)$ are in 
$D$.

To conclude that $(t_1~t_2)$ itself is in $D$, we prove, 
by induction on the construction of $D$, 
that if $D$ is a candidate and 
all the one-step reducts of the term
$(t_1~t_2)$ are in $D$, then $(t_1~t_2)$ is in $D$.
\begin{itemize}
\item If $D = \tilde{\top}$, then as 
all the one-step reducts of the term $(t_1~t_2)$ strongly terminate, 
the term $(t_1~t_2)$ strongly terminates, and
$(t_1~t_2) \in D$.

\item If $D = \tilde{\Pi}(C,S)$, then 
as all the one-step reducts of the term $(t_1~t_2)$ strongly terminate, 
the term $(t_1~t_2)$ strongly terminates. 
If moreover $(t_1~t_2) \lra_{\beta}^* \lambda x:A~v$,
then let $(t_1~t_2) = u_1, u_2, \dots, u_n = \lambda x:A~v$ 
be a reduction sequence from $(t_1~t_2)$ to $\lambda x:A~v$. 
As $(t_1~t_2)$ is an application and $\lambda x:A~v$ is not, $n \geq 2$.
Thus, $(t_1~t_2) \lra_{\beta}^1 u_2 \lra_{\beta}^* \lambda x:A~v$. We have $u_2 \in D$ and 
$u_2 \lra_{\beta}^* \lambda x:A~v$, 
thus for all $w$ in $C$
and $F$ in $S$, $(w/x)v \in F$. 
Thus, $(t_1~t_2) \in \tilde{\Pi}(C,S) = D$.

\item If $D = \bigcap_i D_i$, 
then for all $i$, all the one step reducts of $(t_1~t_2)$ are in 
$D_i$, and, by induction hypothesis $(t_1~t_2) \in D_i$.
Thus, $(t_1~t_2) \in D$.
\end{itemize}
\end{proof}

\subsection{Termination}

Consider a super-consistent theory $\Sigma, {\cal R}$. We want to
prove that $\beta$-reduction terminates in the $\lambda \Pi$-calculus
modulo this theory. 

As usual, we want to associate a candidate $\llbracket A \rrbracket$
to each term $A$ in such a way that if $t$ is a term of type $A$, then
$t \in \llbracket A \rrbracket$. In the $\lambda \Pi$-calculus modulo
theory, the main difficulty is to assign a candidates to terms in such
a way that if $A \equiv B$ then
$\llbracket A \rrbracket = \llbracket B \rrbracket$. For instance, if
we have the rule 
$$P \lra P \Rightarrow P$$
that permits to type all
lambda-terms, including non terminating ones, we should associate, to
the term $P$, a candidate $C$ such that $C = C~\tilde{\Rightarrow}~C$,
but there is no such candidate.  For super-consistent theories, in
contrast, such an assignment exists, as the theory has a model
${\cal M}$ valued in the $\Pi$-algebra
$\langle {\cal C}, \leq, \tilde{\top}, \tilde{\wedge}, {\cal
  P}^{+}({\cal C}), \tilde{\Pi} \rangle$.

Consider this model.

If a term $t$ has type $B$ in some context $\Gamma$, then 
$B$ has type $Type$ in $\Gamma$, 
$B$ has type $Kind$ in $\Gamma$, 
or $B = Kind$. 
Thus, $\llbracket B \rrbracket_{\overline{\phi}}$ is an element of 
${\cal M}_{Type} = {\cal C}$, 
$\llbracket B \rrbracket_{\overline{\phi}}$ is an element of ${\cal M}_{Kind} = 
{\cal C}$, or $\llbracket B \rrbracket_{\overline{\phi}} = \tilde{\top}$. 
In all these cases 
$\llbracket B \rrbracket_{\overline{\phi}}$ is a candidate.

\medskip

\begin{lemma}
Let $\Gamma$ be a context, 
$\overline{\phi} = \phi_1, ..., \phi_n$ is be a sequence of functions 
such that $\phi_i$ maps $x:A$ of $\Gamma$ to an element of 
${\cal D}^i_{A, \phi_1, ..., \phi_{i-1}}$, 
$\sigma$ be a substitution mapping every $x:A$ of $\Gamma$ to an element of 
$\llbracket A \rrbracket_{\overline{\phi}}$ and $t$ a term of type $B$ in $\Gamma$.
Then $\sigma t \in \llbracket B \rrbracket_{\overline{\phi}}$.
\end{lemma}

\begin{proof}
By induction on the structure of the term $t$. 

\begin{itemize}
\item If $t = Type$, then $B = Kind$, 
$\llbracket B \rrbracket_{\overline{\phi}} = \tilde{\top}$
and $\sigma t = Type \in \llbracket B \rrbracket_{\overline{\phi}}$. 

\item If $t = x$ is a variable, then by definition of $\sigma$, 
$\sigma t \in \llbracket B \rrbracket_{\overline{\phi}}$. 

\item If $t = \Pi x:C~D$, then $B = Type$ or $B = Kind$, 
and 
$\llbracket B \rrbracket_{\overline{\phi}} = \tilde{\top}$, 
$\Gamma \vdash C:Type$ and $\Gamma, x:C \vdash D:Type$ or 
$\Gamma, x:C \vdash D:Kind$, by induction hypothesis 
$\sigma C \in \llbracket Type \rrbracket_{\overline{\phi}} = \tilde{\top}$, 
that is $\sigma C$ strongly terminates and
$\sigma D \in \llbracket Type \rrbracket_{\overline{\phi}} = \tilde{\top}$ or
$\sigma D \in \llbracket Kind \rrbracket_{\overline{\phi}} = \tilde{\top}$, 
that is $\sigma D$ strongly terminates.
Thus, $\sigma(\Pi x:C~D) = \Pi x:\sigma C~\sigma D$
strongly terminates also and 
it is an element of $\tilde{\top} = \llbracket B \rrbracket_{\overline{\phi}}$.

\item If $t = \lambda x:C~u$ 
where $u$ has type $D$. Then $B = \Pi x:C~D$ and 
$\llbracket B \rrbracket_{\overline{\phi}}
= 
\llbracket \Pi x:C~D \rrbracket_{\overline{\phi}}
= \tilde{\Pi}(\llbracket C \rrbracket_{\overline{\phi}},
\{\llbracket D \rrbracket_{\overline{\phi}, x = \overline{c}}~|~
\overline{c} \in {\cal D}^1_C \times ... \times {\cal D}^n_{C, \phi_1, ..., \phi_{n-1}}\})$
is the set of terms $s$ such that 
$s$ strongly terminates and if $s$ reduces to $\lambda x:E~s_1$ then
for all $s'$ in 
$\llbracket C \rrbracket_{\overline{\phi}}$ 
and all $\overline{c}$
in ${\cal D}^1_C \times ... \times 
{\cal D}^n_{C, \phi_1, ..., \phi_{n-1}}$, 
$(s'/x)s_1$ is an element of 
$\llbracket D \rrbracket_{\overline{\phi}, x = \overline{c}}$.

We have $\sigma t = \lambda x:\sigma C~\sigma u$, 
consider a reduction sequence issued from this term. This sequence
can only reduce the terms $\sigma C$ and $\sigma u$. 
By induction hypothesis, the term $\sigma C$ is an element of 
$\llbracket Type \rrbracket_{\overline{\phi}} = \tilde{\top}$ 
and the term $\sigma u$ is an
element of $\llbracket D \rrbracket_{\overline{\phi}}$, 
thus the reduction sequence is finite. 

Furthermore, every reduct of $\sigma t$ has the form 
$\lambda x:C'~v$ where $C'$ is a reduct of $\sigma C$ and 
$v$ is a reduct of $\sigma u$. 
Let $w$ be any term of $\llbracket C \rrbracket_{\overline{\phi}}$, 
and $\overline{c}$ be any element of 
${\cal D}^1_C \times ... \times {\cal D}^n_{C, \phi_1, ..., \phi_{n-1}}$,
the term $(w/x)v$ can be obtained by reduction from $((w/x) \circ \sigma)u$. 
By induction hypothesis, the term 
$((w/x) \circ \sigma)u$ is an element of 
$\llbracket D \rrbracket_{\overline{\phi}, x = \overline{c}}$. 
Hence, by Lemma \ref{closurepi}
the term $(w/x)v$ is an element of 
$\llbracket D \rrbracket_{\overline{\phi}, x = \overline{c}}$. 
Therefore, the term $\sigma \lambda x~u$ is an
element of $\llbracket B \rrbracket_{\overline{\phi}}$. 

\item
If the term $t$ has the form $(u_1~u_2)$ then $u_1$ is a term of type 
$\Pi x:C~D$, $u_2$ a term of type $C$ and $B = (u_2/x)D$.
We have $\sigma t = (\sigma u_1~\sigma u_2)$,
and by induction hypothesis
$\sigma u_1 \in \llbracket \Pi x:C~D \rrbracket_{\overline{\phi}} = 
\tilde{\Pi}(\llbracket C \rrbracket_{\overline{\phi}}, 
\{\llbracket D \rrbracket_{\overline{\phi}, x = \overline{c}}~|~\overline{c} \in 
{\cal D}^1_C \times ... \times {\cal D}^n_{C, \phi_1, ..., \phi_{n-1}}\})$
and $\sigma u_2 \in \llbracket C \rrbracket_{\overline{\phi}}$. 
By Lemma \ref{application}, 
$(\sigma u_1~\sigma u_2) \in 
\llbracket D \rrbracket_{\overline{\phi}, x = 
{\cal D}^1_{u_2}, ..., {\cal D}^n_{u_2, \phi_1, ..., \phi_{n-1}}}
= \llbracket (u_2/x)D \rrbracket_{\overline{\phi}}
= \llbracket B \rrbracket_{\overline{\phi}}$.
\end{itemize}
\end{proof}

\begin{theorem}
Let $t$ be a term well-typed in a context $\Gamma$.
Then $t$ strongly terminates.
\end{theorem}

\begin{proof}
Let $B$ be the type of $t$ in $\Gamma$, 
let 
$\overline{\phi} = \phi_1, ..., \phi_n$ is be a sequence of functions 
such that $\phi_i$ maps $x:A$ of $\Gamma$ 
to an element of ${\cal D}^i_{A, \phi_1, ..., \phi_{i-1}}$, 
$\sigma$ be the substitution mapping every $x:A$ of $\Gamma$ to itself.
Note that, by Lemma \ref{Varpi}, 
this variable is an element of 
$\llbracket A \rrbracket_{\overline{\phi}}$. 
Then $t = \sigma t \in \llbracket B \rrbracket_{\overline{\phi}}$.
Hence it strongly terminates.
\end{proof}

\subsection{Termination of the $\beta {\cal R}$-reduction}
\label{sec:termbr}

We finally prove the termination of the $\beta {\cal R}$-reduction for 
Simple type theory without or with a parametric quantifier
and for the Calculus of constructions.
The rules ${\cal R}$ of Simple type theory are
$$\varepsilon~(\Rightarrow~x~y) \lra (\varepsilon~x) \ra (\varepsilon~y)$$
$$\varepsilon~(\fa_A~x) \lra \Pi z:A~(\varepsilon~(x~z))$$
This set ${\cal R}$ of rewrite rules terminates, as each reduction step
reduces the number of symbols
$\Rightarrow$ and $\fa_A$ in the term.
Then, ${\cal R}$-reduction can create $\beta$-redices, but only 
$\beta$-redices on the form $((\lambda x:A~t)~z)$ where $z$ is a 
variable. Thus, any term can be weakly $\beta {\cal R}$-reduced by 
$\beta$-reducing it first, then ${\cal R}$-reducing it, then 
$\beta$-reducing the trivial $\beta$-redices created by the 
${\cal R}$-reduction.

A similar argument applies to Simple type theory with a parametric quantifier
and to the Calculus of constructions.

\section*{Acknowledgements}

The author wants to thank Fr\'ed\'eric Blanqui for very helpful remarks 
on a previous version of this paper.

\bibliography{superpi}

\begin{thebibliography}{10}

\bibitem{expressing}
A.~Assaf, G.~Burel, R.~Cauderlier, D.~Delahaye, G.~Dowek, C.~Dubois,
  F.~Gilbert, P.~Halmagrand, O.~Hermant, and R.~Saillard.
\newblock Dedukti: a logical framework based on the lambda-{P}i-calculus modulo
  theory.
\newblock {\tt http://www.lsv.ens-cachan.fr/\~{}dowek/Publi/} {\tt
  expressing.pdf}, 2016.

\bibitem{Jouannaud}
A.~Assaf, G.~Dowek, J.-P. Jouannaud, and J.~Liu.
\newblock Untyped confluence in dependent type theories.
\newblock Submitted to publication, 2017.

\bibitem{Baueretal}
A.~Bauer, G.~Gilbert, P.~Haselwarter, M.~Pretnar, and Ch.~A. Stone.
\newblock Design and implementation of the andromeda proof assistant.
\newblock Types, 2016.

\bibitem{Blanqui}
F.~Blanqui.
\newblock Definitions by rewriting in the calculus of constructions.
\newblock {\em Mathematical Structures in Computer Science}, 15(1):37--92,
  2005.

\bibitem{BHH}
A.~Brunel, O.~Hermant, and C.~Houtmann.
\newblock Orthogonality and boolean algebras for deduction modulo.
\newblock In L.~Ong, editor, {\em Typed Lambda Calculus and Applications},
  volume 6690 of {\em Lecture Notes in Computer Science}, pages 76--90.
  Springer-Verlag, 2011.

\bibitem{CirsteaLiquoriWack}
H.~Cirstea, L.~Liquori, and B.~Wack.
\newblock Rewriting calculus with fixpoints: Untyped and first-order systems.
\newblock In {\em Types}, volume 3085 of {\em Lectures Notes in Computer
  Science}. Springer-Verlag, 2003.

\bibitem{CoquandHuet}
T.~Coquand and G.~Huet.
\newblock The calculus of constructions.
\newblock {\em Information and Computation}, pages 95--120, 1988.

\bibitem{CD}
D.~Cousineau and G.~Dowek.
\newblock Embedding pure type systems in the lambda-pi-calculus modulo.
\newblock In S.~Ronchi~Della Rocca, editor, {\em Typed lambda calculi and
  applications}, volume 4583 of {\em Lecture Notes in Computer Science}, pages
  102--117. Springer-Verlag, 2007.

\bibitem{TVA}
G.~Dowek.
\newblock Truth values algebras and proof normalization.
\newblock In Th. Altenkirch and C.~McBride, editors, {\em Types for proofs and
  programs}, volume 4502 of {\em Lecture Notes in Computer Science}, pages
  110--124. Springer-Verlag, 2007.

\bibitem{DHKHOL}
G.~Dowek, Th. Hardin, and C.~Kirchner.
\newblock Hol-lambda-sigma: an intentional first-order expression of
  higher-order logic.
\newblock {\em Mathematical Structures in Computer Science}, 11:1--25, 2001.

\bibitem{DHK}
G.~Dowek, Th. Hardin, and C.~Kirchner.
\newblock Theorem proving modulo.
\newblock {\em Journal of Automated Reasoning}, 31:33--72, 2003.

\bibitem{DW}
G.~Dowek and B.~Werner.
\newblock Proof normalization modulo.
\newblock {\em \em The Journal of Symbolic Logic}, 68(4):1289--1316, 2003.

\bibitem{FosterStruth}
S.~Foster and G.~Struth.
\newblock Integrating an automated theorem prover into agda.
\newblock In M.~Bobaru, K.~Havelund, G.J. Holzmann, and R.~Joshi, editors, {\em
  NASA Formal Methods}, volume 6617 of {\em Lecture Notes in Computer Science}.
  Springer-Verlag, 2011.

\bibitem{Geuvers}
H.~Geuvers.
\newblock A short and flexible proof of strong normalization for the calculus
  of constructions.
\newblock In P.~Dybjer, , B.~Nordstr{\"o}m, and J.~Smith, editors, {\em Types
  for Proofs and Programs}, volume 996 of {\em Lecture Notes in Computer
  Science}, pages 14--38. Springer-Verlag, 1995.

\bibitem{Girard}
J.Y. Girard.
\newblock {\em Interpr\'etation Fonctionnelle et \'Elimination des Coupures
  dans l'Arithm\'etique d'Ordre Sup\'erieur}.
\newblock PhD thesis, Universit\'e de Paris VII, 1972.

\bibitem{HHP}
R.~Harper, F.~Honsell, and G.~Plotkin.
\newblock A framework for defining logics.
\newblock {\em Journal of the ACM}, 40(1):143--184, 1993.

\bibitem{MartinLof}
P.~Martin-L\"of.
\newblock {\em Intuitionistic Type Theory}.
\newblock Bibliopolis, 1984.

\bibitem{MelliesWerner}
P.-A. Melli{\`e}s and B.~Werner.
\newblock A generic normalisation proof for pure type systems.
\newblock In E.~Gim{\'e}nez and Ch. Paulin-Mohring, editors, {\em Types for
  Proofs and Programs}, volume 1512 of {\em Lecture Notes in Computer Science},
  pages 254--276. Springer-Verlag, 1998.

\bibitem{MiquelWerner}
A.~Miquel and B.~Werner.
\newblock The not so simple proof-irrelevant model of {CC}.
\newblock In H.~Geuvers and F.~Wiedijk, editors, {\em Types for Proofs and
  Programs}, pages 240--258. Springer-Verlag, 2003.

\bibitem{NguyenKirchnerKirchner}
Q.H. Nguyen, C.~Kirchner, and H.~Kirchner.
\newblock External rewriting for skeptical proof assistants.
\newblock {\em Journal of Automated Reasoning}, 29(309), 2002.

\bibitem{NPS}
B.~Nordstr\"om, K.~Petersson, and J.M. Smith.
\newblock Martin-l\"of's type theory.
\newblock In S.~Abramsky, D.~Gabbay, and T.~Maibaum, editors, {\em Handbook of
  Logic in Computer Science}, pages 1--37. Clarendon Press, 2000.

\bibitem{Parigot}
M.~Parigot.
\newblock Proofs of strong normalization for second order classical natural
  deduction.
\newblock In {\em Logic in Computer Science}, pages 39--46, 1993.

\end{thebibliography}

\cleardoublepage

\section{Appendix: Super-consistency of the Calculus of constructions}

Let ${\cal B} = \langle {\cal B}, \tilde{\top}, \tilde{\wedge}, 
{\cal P}^+({\cal B}), \tilde{\Pi} \rangle$ 
be a full $\Pi$-algebra. 
Let $\{e\}$ be an arbitrary one-element set.  Let ${\cal U}$ 
and ${\cal V}$ defined as in Section 
\ref{sttp}. 

Note that ${\cal U}$ does not need to be
closed by dependent function space. This can be compared with the fact
that all terms that can be typed in the Calculus of constructions can
be typed in the system $F\omega$.

\medskip
\begin{definition}
The interpretation function ${\cal N}$ is defined as follows

\begin{itemize}
\item ${\cal N}_{Type} = {\cal N}_{Kind} = {\cal V}$,
\item ${\cal N}_{\Pi x:C~D}$ is the set ${\cal F}({\cal N}_C,{\cal N}_{D})$, 
except if ${\cal N}_{D} = \{e\}$, in which case ${\cal N}_{\Pi x:C~D} = \{e\}$,
\item ${\cal N}_{type} = {\cal U}$,
\item ${\cal N}_{o} = 
{\cal N}_{\dot{\Pi}_{KK}} = 
{\cal N}_{\dot{\Pi}_{TT}} = 
{\cal N}_{\dot{\Pi}_{KT}} = 
{\cal N}_{\dot{\Pi}_{TK}} = 
{\cal N}_{\eta} = 
{\cal N}_{\varepsilon} = \{e\}$, 
\item ${\cal N}_{x} = \{e\}$, 
\item ${\cal N}_{\lambda x:C~t} = {\cal N}_{t}$,
\item ${\cal N}_{(t~u)} = {\cal N}_{t}$.
\end{itemize}
\end{definition}

\medskip

We first prove the two following lemmas.

\begin{lemma}\label{lem1cc}
If the term $t$ is an object, then 
$${\cal N}_t = \{e\}$$
\end{lemma}

\begin{proof}
By induction on the structure of the term $t$. 
The term $t$ is neither $Kind$, $Type$, nor $type$. It is not a product. 
If it has the form $\lambda x:C~t'$, then $t'$ is an object.
If it has the form $(t'~t'')$, then $t'$ is an object.
\end{proof}

\begin{lemma}\label{lem2cc}
If $u$ is an object, then 
$${\cal N}_{(u/x)t} = {\cal N}_t$$
\end{lemma}

\begin{proof}
By induction on the structure of the term $t$. 
If $t = x$ then,
by Lemma \ref{lem1cc}
$${\cal N}_{(u/x)t} = {\cal N}_u = \{e\} = {\cal N}_t$$ If $t$ is
$Kind$, $Type$, a constant, or a variable different from $x$, then $x$
does not occur in $t$.  If it is a product, an abstraction, or an
application, we use the induction hypothesis.
\end{proof}

\begin{lemma}[Validity of the congruence]
If $t \equiv_{\beta {\cal R}} u$ then 
$${\cal N}_t = {\cal N}_u$$
\end{lemma}

\begin{proof}
If $t = ((\lambda x:C~t')~u')$, then $u'$ is an object, then by
Lemma \ref{lem2cc}
$${\cal N}_{((\lambda x:C~t')~u')} = {\cal N}_{t'} = {\cal N}_{(u'/x)t'}$$
Then, as for all $v$, ${\cal N}_{(\eta~v)} = {\cal N}_{\eta} = \{e\}$, 
for all $w$, ${\cal N}_{(\varepsilon~w)} = {\cal N}_{\varepsilon} 
= \{e\}$, 
and if ${\cal N}_D = \{e\}$, then ${\cal N}_{\Pi x:C~D} = \{e\}$, 
we have 
$${\cal N}_{(\eta~(\dot{\Pi}_{KK}~C~D))} = \{e\} 
= {\cal N}_{\Pi x:(\eta~C)~(\eta~(D~x))}$$
$${\cal N}_{(\varepsilon~(\dot{\Pi}_{TT}~C~D))} = \{e\} 
= {\cal N}_{\Pi x:(\varepsilon~C)~(\varepsilon~(D~x))}$$
$${\cal N}_{(\varepsilon~(\dot{\Pi}_{KT}~C~D))} = \{e\} 
= {\cal N}_{\Pi x:(\eta~C)~(\varepsilon~(D~x))}$$
$${\cal N}_{(\eta~(\dot{\Pi}_{TK}~C~D))} = \{e\} 
= {\cal N}_{\Pi x:(\varepsilon~C)~(\eta~(D~x))}$$
We prove, by induction on $t$, that if 
$t \lra^1_{\beta {\cal R}} u$ then ${\cal N}_t = {\cal N}_u$
and we conclude with a simple induction on the structure of a reduction 
of $t$ and $u$ to a common term. 
\end{proof}

\begin{definition}
The interpretation function ${\cal M}$ is defined as follows

\begin{itemize}
\item 
${\cal M}_{Kind,\psi} = {\cal M}_{Type,\psi}  = {\cal B}$,

\item ${\cal M}_{\Pi x:C~D, \psi}$ is the set of functions $f$ mapping
$\langle c', c\rangle$ in 
${\cal N}_C \times {\cal M}_{C,\psi}$ to an element of 
${\cal M}_{D,(\psi, x = c')}$, except if for all $c'$ in 
${\cal N}_C$, ${\cal M}_{D,(\psi, x = c')} = \{e\}$, in which case 
${\cal M}_{\Pi x:C~D, \psi} = \{e\}$,

\item ${\cal M}_{type,\psi}  = {\cal B}$,

\item ${\cal M}_{\eta,\psi}$ is the function of ${\cal F}({\cal U},{\cal V})$ 
mapping $S$ to $S$,

\item ${\cal M}_{\varepsilon,\psi}$ is the function of
${\cal F}(\{e\}, {\cal V})$, mapping $e$ to $\{e\}$, 

\item ${\cal M}_{o,\psi} = {\cal B}$,

\item ${\cal M}_{\dot{\Pi}_{KK}}$ is the function 
mapping $S$ in ${\cal U}$ and $h$ in ${\cal F}(\{e\}, {\cal U})$ 
to the set ${\cal F}(\{e\} \times S,(h~e))$, 
except if $(h~e) = \{e\}$ in which case it maps $S$ and $h$ to $\{e\}$, 

\item ${\cal M}_{\dot{\Pi}_{TT}} = e$, 

\item ${\cal M}_{\dot{\Pi}_{KT}} = e$, 

\item ${\cal M}_{\dot{\Pi}_{TK}}$ is the function 
mapping $e$ and $h$ in ${\cal F}(\{e\}, {\cal U})$ 
to the set ${\cal F}(\{e\}\times\{e\}, (h~e))$, except if $(h~e) = \{e\}$ 
in which case it maps $e$ and $h$ to $\{e\}$, 

\item ${\cal M}_{x,\psi} = \psi x$,

\item ${\cal M}_{\lambda x:C~t,\psi}$ is the function mapping $c$ in ${\cal N}_C$
to ${\cal M}_{t, (\psi, x = c)}$, except if for all $c$ in ${\cal N}_C$,
${\cal M}_{t, (\psi, x = c)} = e$ in which case 
${\cal M}_{\lambda x:C~t,\psi}  = e$,

\item ${\cal M}_{(t~u),\psi} = {\cal M}_{t,\psi}~{\cal M}_{u,\psi}$,
except if ${\cal M}_{t,\psi} = e$ in which case 
${\cal M}_{(t~u),\psi} = e$.
\end{itemize}
\end{definition}

\medskip

\begin{lemma}
If $\Gamma \vdash C:Type$, then 
$${\cal N}_C \in {\cal V}$$
\end{lemma}

\begin{proof}
By induction on the structure of the term $C$. As this term has type
$Type$, it is neither $Kind$ nor $Type$.
\end{proof}

\begin{lemma}[Well-typedness]
If $\Gamma \vdash t:B$ 
and $\psi$ is a function mapping the variables
$x:A$ of $\Gamma$ to elements of ${\cal N}_A$, then 
$${\cal M}_{t,\psi} \in {\cal N}_B$$
\end{lemma}

\begin{proof}
We check each case of the definition of ${\cal M}$.
\end{proof}

\begin{lemma}[Substitution]
For all $t$, $u$ and $\psi$
$${\cal M}_{(u/x)t,\psi} = {\cal M}_{t, (\psi, x = {\cal M}_u)}$$
\end{lemma}

\begin{proof}
By induction on the structure of the term $t$.
\end{proof}

\begin{lemma}[Validity of the congruence]
If $t \equiv_{\beta {\cal R}} u$ then 
$${\cal M}_{t,\psi} = {\cal M}_{u,\psi}$$
\end{lemma}

\begin{proof}
If $t = ((\lambda x:C~t')~u')$, then if for all $c$ in ${\cal N}_C$ 
${\cal M}_{t',(\psi, x = c)} = e$ then 
$${\cal M}_{((\lambda x:C~t')~u'),\psi} = e = 
{\cal M}_{t',(\psi, x = {\cal M}_{u',\psi})} = 
{\cal M}_{(u'/x)t',\psi}$$
Otherwise 
$${\cal M}_{((\lambda x:C~t')~u'),\psi} = 
{\cal M}_{t', (\psi, x = {\cal M}_{u',\psi})} = {\cal M}_{(u'/x)t',\psi}$$
The set 
${\cal M}_{(\eta~(\dot{\Pi}_{KK}~C~D)),\psi}$ 
is the set ${\cal F}(\{e\} \times {\cal M}_{C,\psi},({\cal M}_{D,\psi}~e))$, except if 
$({\cal M}_{D,\psi}~e) = \{e\}$ in which case 
${\cal M}_{(\eta~(\dot{\Pi}_{KK}~C~D)),\psi} = 
\{e\}$. The set ${\cal M}_{\Pi x:(\eta~C)~(\eta~(D~x)),\psi}$
is this same set. Thus
$${\cal M}_{(\eta~(\dot{\Pi}_{KK}~C~D)),\psi} 
= {\cal M}_{\Pi x:(\eta~C)~(\eta~(D~x)),\psi}$$
We have 
$${\cal M}_{(\varepsilon~(\dot{\Pi}_{TT}~C~D))} = \{e\}
= {\cal M}_{\Pi x:(\varepsilon~C)~(\varepsilon~(D~x))}$$
and
$${\cal M}_{(\varepsilon~(\dot{\Pi}_{KT}~C~D))} = \{e\} 
= {\cal M}_{\Pi x:(\varepsilon~C)~(\varepsilon~(D~x))}$$
The set 
${\cal M}_{(\eta~(\dot{\Pi}_{TK}~C~D)),\psi}$ 
is the set ${\cal F}(\{e\} \times \{e\},({\cal M}_{D,\psi}~e))$, except if 
$({\cal M}_{D,\psi}~e) = \{e\}$ in which case 
${\cal M}_{(\eta~(\dot{\Pi}_{TK}~C~D)),\psi} = 
\{e\}$. The set ${\cal M}_{\Pi x:(\varepsilon~C)~(\eta~(D~x)),\psi}$
is this same set. Thus
$${\cal M}_{(\eta~(\dot{\Pi}_{TK}~C~D)),\psi} 
= {\cal M}_{\Pi x:(\varepsilon~C)~(\eta~(D~x)),\psi}$$
We prove, by induction on $t$, that if $t \lra^1_{\beta {\cal R}} u$
then ${\cal M}_{t,\psi} = {\cal M}_{u,\psi}$ 
and we conclude with a simple induction on the structure of a reduction 
of $t$ and $u$ to a common term. 
\end{proof}

\begin{definition}
The interpretation function $\llbracket . \rrbracket$ is defined as follows

\begin{itemize}
\item 
$\llbracket Kind\rrbracket_{\psi,\phi} = \llbracket Type\rrbracket_{\psi,\phi} =
\tilde{\top}$,

\item $\llbracket \Pi x:C~D \rrbracket_{\psi,\phi} = 
\tilde{\Pi}(\llbracket C \rrbracket_{\psi,\phi}, 
\{\llbracket D \rrbracket_{(\psi, x = c'),(\phi, x = c)}~|~c' \in {\cal N}_C, c \in 
{\cal M}_{C, \psi}\})$,

\item $\llbracket type \rrbracket_{\psi,\phi} = \tilde{\top}$,

\item $\llbracket o \rrbracket_{\psi,\phi} = \tilde{\top}$,

\item $\llbracket \dot{\Pi}_{KK} \rrbracket_{\psi,\phi}$ 
is the function mapping 
$\langle S, a \rangle$ in ${\cal U} \times {\cal B}$, 
$\langle f,g\rangle$ in 
${\cal F}(\{e\}, {\cal U}) \times {\cal F}(\{e\} \times S, {\cal B})$ to
$\tilde{\Pi}(a,\{(g~\langle e, s\rangle)~|~s \in S\})$,

\item $\llbracket \dot{\Pi}_{TT} \rrbracket_{\psi,\phi}$ 
is the function mapping 
$\langle e, a \rangle$ in $\{e\} \times {\cal B}$, 
and $\langle e, g\rangle$ in $\{e\} \times {\cal F}(\{e\} \times \{e\}, 
{\cal B})$ to $\tilde{\Pi}(a,\{(g~\langle e, e \rangle)\})$,

\item $\llbracket \dot{\Pi}_{KT} \rrbracket_{\psi,\phi}$ 
is the function mapping 
$\langle S,a \rangle$ in ${\cal U} \times {\cal B}$, 
and $\langle e, g \rangle$ in $\{e\} \times {\cal F}(\{e\} \times S, {\cal B})$ 
to $\tilde{\Pi}(a,\{(g~\langle e, s \rangle)~|~s \in S\})$,

\item $\llbracket \dot{\Pi}_{TK} \rrbracket_{\psi,\phi}$
is the function mapping 
$\langle e, a \rangle$ in $\{e\} \times {\cal B}$, 
and $\langle f, g \rangle$ in ${\cal F}(\{e\},{\cal U}) \times
{\cal F}(\{e\} \times \{e\}, {\cal B})$ to
$\tilde{\Pi}(a,\{(g~\langle e, e \rangle)\})$,

\item $\llbracket \eta \rrbracket_{\psi,\phi}$ is 
the function from ${\cal U} \times 
{\cal B}$ to ${\cal B}$, mapping $\langle S, a \rangle$ to $a$, 

\item $\llbracket \varepsilon \rrbracket_{\psi,\phi}$
is the function from 
$\{e\} \times {\cal B}$ to ${\cal B}$,
mapping $\langle e, a \rangle$ to $a$, 

\item $\llbracket x\rrbracket_{\psi,\phi} = \phi x$, 

\item $\llbracket \lambda x:C~t\rrbracket_{\psi,\phi}$ is the function 
mapping
$\langle c', c \rangle$ in ${\cal N}_C \times {\cal M}_{C, \psi}$ to 
$\llbracket t \rrbracket_{(\psi, x = c'),(\phi, x = c)}$,
except if for all 
$\langle c', c \rangle$ in ${\cal N}_C \times {\cal M}_{C,\psi}$, 
$\llbracket t \rrbracket_{(\psi, x= c'),(\phi, x = c)} = e$, in which 
case $\llbracket \lambda x:C~t\rrbracket_{\psi,\phi} = e$,

\item $\llbracket (t~u)\rrbracket_{\psi,\phi} = 
\llbracket t \rrbracket_{\psi,\phi}~\langle {\cal M}_{u, \psi}, 
\llbracket u \rrbracket_{\psi,\phi} \rangle$, except 
if $\llbracket t \rrbracket_{\psi,\phi} = e$, in which case 
$\llbracket (t~u) \rrbracket_{\psi,\phi} = e$.
\end{itemize}
\end{definition}

\medskip

\begin{lemma}[Well-typedness]
If $\Gamma \vdash t:B$, 
$\psi$ is a function mapping variables $x:A$ of $\Gamma$ to 
elements of ${\cal N}_A$, and $\phi$ is a function mapping variables 
$x:A$ of $\Gamma$ to elements of ${\cal M}_{A,\psi}$, 
then 
$$\llbracket t \rrbracket_{\psi,\phi} \in {\cal M}_{B,\psi}$$
\end{lemma}

\begin{proof}
We check each case of the definition of $\llbracket . \rrbracket$.
\end{proof}

\medskip 

\begin{lemma}[Substitution]
For all $t$, $u$, $\psi$, and $\phi$
$$\llbracket (u/x)t \rrbracket_{\psi, \phi} = \llbracket t \rrbracket_{
(\psi, x = {\cal M}_{u,\psi}), 
(\phi, x = \llbracket u \rrbracket_{\psi,\phi}})$$
\end{lemma}

\begin{proof}
By induction on the structure of the term $t$.
\end{proof}

\begin{lemma}[Validity of the congruence]
If $t \equiv_{\beta {\cal R}} u$ then 
$$\llbracket t \rrbracket_{\psi,\phi} = \llbracket u \rrbracket_{\psi,\phi}$$
\end{lemma}

\begin{proof}
If $t = ((\lambda x:C~t')~u')$, then 
if for all $c'$ in ${\cal N}_C$ and $c$ in ${\cal M}_{C,\psi}$, we
have $\llbracket t' \rrbracket_{(\psi, x = c'),(\phi, x = c)} = e$
then 
$$\llbracket ((\lambda x:C~t')~u') \rrbracket_{\psi,\phi} = 
e = 
\llbracket t' \rrbracket_{(\psi, x = {\cal M}_{u', \psi}),(\phi, x = 
\llbracket u' \rrbracket_{\psi,\phi})}
= 
\llbracket (u'/x)t' \rrbracket_{\psi,\phi}$$
Otherwise
$$\llbracket ((\lambda x:C~t')~u') \rrbracket_{\psi,\phi}
= \llbracket t' \rrbracket_{(\psi, x = {\cal M}_{u', \psi}),(\phi, x = 
                                      \llbracket u' \rrbracket_{\psi,\phi})}
= \llbracket (u'/x)t' \rrbracket_{\psi,\phi}$$
We have
$$\llbracket (\eta~(\dot{\Pi}_{KK}~C~D)) \rrbracket_{\psi,\phi}
= \tilde{\Pi}(\llbracket C \rrbracket_{\psi,\phi}, \{(\llbracket D \rrbracket_{\psi,\phi}~\langle e, s \rangle)~|~s \in {\cal M}_{C,\psi}\})
= \llbracket \Pi y:(\eta~C)~(\eta~(D~y)) \rrbracket_{\psi,\phi}$$
$$\llbracket (\varepsilon~(\dot{\Pi}_{TT}~C~D))
\rrbracket_{\psi,\phi} = 
\tilde{\Pi}(\llbracket C \rrbracket_{\psi,\phi}, \{(\llbracket D \rrbracket_{\psi,\phi}~\langle e, e \rangle)\})
= \llbracket \Pi y:(\varepsilon~C)~(\varepsilon~(D~y)) 
\rrbracket_{\psi,\phi}$$
$$\llbracket (\varepsilon~(\dot{\Pi}_{KT}~C~D)) \rrbracket_{\psi,\phi}
= \tilde{\Pi}(\llbracket C \rrbracket_{\psi,\phi}, \{(\llbracket D \rrbracket_{\psi,\phi}~\langle e, s \rangle)~|~s \in {\cal M}_{C,\psi}\})
= \llbracket \Pi y:(\eta~C)~(\varepsilon~(D~y)) \rrbracket_{\psi,\phi}$$
$$\llbracket (\eta~(\dot{\Pi}_{TK}~C~D)) \rrbracket_{\psi,\phi}
= \tilde{\Pi}(\llbracket C \rrbracket_{\psi,\phi}, 
\{(\llbracket D \rrbracket_{\psi,\phi}\langle e, e \rangle)\})
= \llbracket \Pi y:(\varepsilon~C)~(\eta~(D~y)) \rrbracket_{\psi,\phi}$$
We prove, by induction on $t$, that if 
$t \lra^1_{\beta {\cal R}} u$ then $\llbracket t \rrbracket_{\psi,\phi} = \llbracket u 
\rrbracket_{\psi,\phi}$
and we conclude with a simple induction on the structure of a reduction 
of $t$ and $u$ to a common term. 
\end{proof}

\end{document}